\documentclass[useAMS,usenatbib]{mn2e}
\usepackage{psfig}
\usepackage{amssymb}
\newcommand{\HI}{H\,{\sc i}}

\newcommand{\kms}{km\,s$^{-1}$}
\newcommand{\kkms}{km\,s$^{-1}$}

\newcommand{\MHI}{$M_{\rm HI}$}

\newcommand{\Msun}{$M_{\odot}$}

\title[Gaseous Tidal Debris found in the NGC 3783 Group]{Gaseous Tidal Debris found in the NGC 3783 Group\thanks{The observations were obtained with the Australia
       Telescope which is funded by the Commonwealth of Australia for
       operations as a National Facility managed by CSIRO.}}

\author[V. A. Kilborn et al.]
       {Virginia A. Kilborn$^{1,2}$, 
        Duncan A. Forbes$^{1}$, 
        B\"arbel S. Koribalski$^2$,
     	Sarah Brough$^{1}$,\newauthor 
	Katie Kern$^{1,2}$\\
        $^1$Centre for Astrophysics \& Supercomputing, Swinburne University, Hawthorn, VIC 3122, Australia\\
        $^2$Australia Telescope National Facility, CSIRO, 
            P.O. Box 76, Epping, NSW 1710, Australia\\}

\begin{document}

\date{Accepted; Received }

\pagerange{\pageref{firstpage}--\pageref{lastpage}} \pubyear{2002}

\maketitle

\label{firstpage}

\begin{abstract}

We have conducted wide-field \HI\ mapping of a $\sim 5.5\degr \times
5.5\degr$ region surrounding the NGC 3783 galaxy group, to an \HI\
mass limit of $\sim 4\times 10^8$\Msun.  The observations were made
using the multibeam system on the Parkes 64-m radiotelescope, as part
of the Galaxy Evolution Multiwavelength Study (GEMS).  We find twelve
\HI\ detections in our Parkes data, four more than catalogued in
HIPASS. We find two new group members, and discover an isolated region
of \HI\ gas with an \HI\ mass of $\sim 4 \times 10^8$\Msun, without a
visible corresponding optical counterpart. We discuss the likelihood
of this \HI\ region being a low surface brightness galaxy, primordial
gas, or a remnant of tidal debris. For the NGC 3783 group we derive a
mean recession velocity of 2903$\pm$26 \kms, and a velocity dispersion
of 190$\pm24$ \kms. The galaxy NGC 3783 is the nearest galaxy to the
luminosity weighted centre of the group, and is at the group mean
velocity. From the X-ray and dynamical state of this galaxy group,
this group appears to be in the early stages of its evolution.

\end{abstract}

\begin{keywords}
galaxies: cluster: general -- galaxies: evolution -- galaxies: irregular
\end{keywords}

\begin{figure*} 
\begin{tabular}{c}
\mbox{\psfig{file=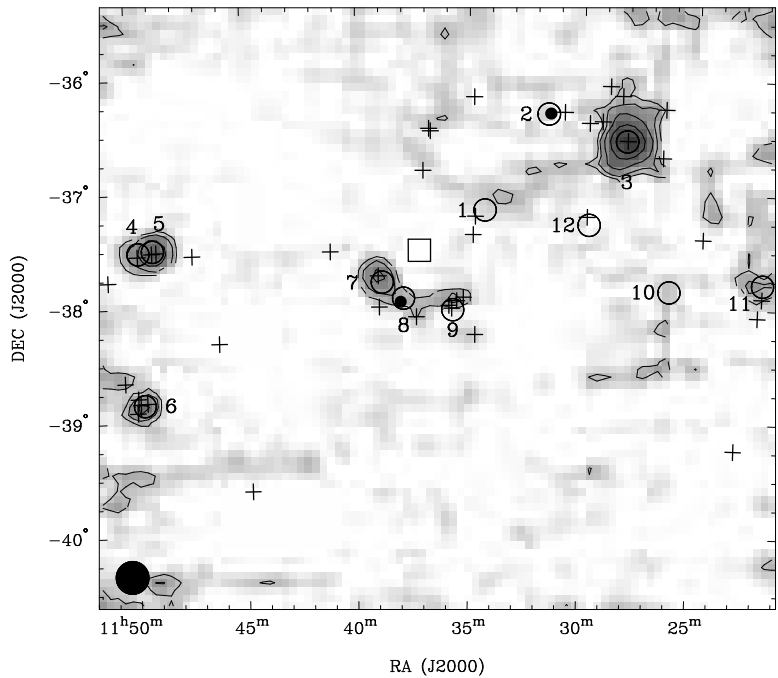,height=10cm,angle=-90}}
\end{tabular}
\caption{Velocity integrated total \HI\ map of the NGC 3783 group. The
contour levels are 2, 4, 8, 16, 32, 64 Jy \kms. The large open circles mark
the HI detections (numbered 1-12), the crosses mark galaxies with known
redshifts in the region from NED and 6dFGS DR2 with velocities between
2500--3500 \kms. The filled circles mark the positions of the
previously uncatalogued \HI\ detections GEMS\_N3783\_2 and
GEMS\_N3783\_8 from the ATCA. The open square marks the luminosity-weighted centre
of the group. Galaxy \#7 is NGC 3783 itself, which lies at the centre
of an extended X-ray halo. The Parkes beam is shown in the bottom left
corner.}
\label{fig:hi_mom0}
\end{figure*}

\section{Introduction}

The physical processes operating in galaxy groups are likely to be
somewhat different to those operating in galaxy clusters. For example,
the intragroup medium is significantly less dense than the
intracluster medium and galaxy motions slower -- suggesting that ram
pressure stripping is less effective in groups than in clusters
(Mulchaey \& Zabludoff 1998). On the other hand, as relative galaxy
motions are smaller, gravitational interactions will be more effective
and outright mergers more common in groups (Zabludoff \& Mulchaey
1998).

Neutral hydrogen (\HI) observations are a particularly useful method
of probing the relative importance of ram pressure stripping and tidal
interactions. The \HI\ is typically more extended than the optical
disk of a galaxy, and thus is more sensitive to tidal interactions
than the stars (e.g. Haynes, Giovanelli \& Chincarini 1984), and ram
pressure stripping affects the gas rather than the stars in a
galaxy. In addition, \HI\ surveys can detect new, optically faint but
gas-rich galaxies, previously unidentified in optical surveys
(e.g. Kilborn et al. 2005; Banks et al. 1999). However, systematic,
optically unbiased \HI\ surveys of loose galaxy groups are few. These
\HI\ surveys of groups have found galaxies with extended HI envelopes
(Haynes 1981), some galaxies with significant \HI\ deficiencies
(Kilborn et al. 2005; Omar \& Dwarakanath 2005) and that there are no
intergalactic \HI\ clouds in groups similar to the Local Group (Pisano
et al. 2004; Zwaan 2001).

Tidal interactions between gas-rich galaxies (e.g. Barnes \& Hernquist
1986; Toomre \& Toomre 1972) are known to produce \HI\ bridges
(e.g. Koribalski \& Manthey 2005; Koribalski \& Dickey 2004; Yun, Lo
\& Ho 1994; Li \& Seaquist 1994), tails (e.g. Hibbard et al. 2001b;
Hibbard \& Yun 1999) or even `clouds' that are spatially distinct from
the host galaxy (e.g. Bekki, Koribalski \& Kilborn 2005a; Bekki et
al. 2005b; Ryder et al. 2001; Schneider 1985). Such \HI\ clouds, with
no obvious optical counterpart, have been detected in clusters
(Oosterloo \& van Gorkom 2005; Davies et al. 2004; Giovanelli \&
Haynes 1989) but their origin is not always immediately obvious (Bekki
et al. 2005a; Minchin et al. 2005; Djorgovski 1990). Determining the
origin of \HI\ `clouds' is vital. For example, cosmological n-body
simulations of the Local Group predict many more dark matter halos,
than galaxies are optically observed in the Local Group (Moore et
al. 1999; Klypin et al. 1999). Could these halos be detected in an
\HI\ survey?  Despite deep \HI\ surveys of groups (Pisano et al. 2004;
Zwaan 2001), and the HIPASS southern sky \HI\ survey (Meyer et
al. 2004), we are yet to identify any \HI\ sources that are
primordial, i.e. are not associated with a star forming galaxy.

%\HI\ observations of galaxies in clusters have revealed that some disk
%galaxies contain much less \HI\ than expected (Solanes et al. 2001;
%Cayatte et al. 1990; Giovanelli \& Haynes 1985), and are commonly
%labeled `\HI\ deficient'. The source of the cold gas removal is
%usually attributed to ram pressure stripping (e.g. Abadi, Moore \&
%Bower 1999; Vollmer et al. 2001), however it is clear that tidal
%interactions play a part in some instances (e.g. Toomre \& Toomre
%1972; Vollmer 2003; Bekki et al. 2005b). Studies of galaxies in
%compact groups (with space densities similar to the cores of clusters,
%but with much lower velocity dispersions) have yielded contradictory
%conclusions as to whether they are \HI\ deficient or not
%(Verdes-Montenegro et al. 2001; Stevens et al. 2004). Similar studies
%of the HI content of galaxies in loose groups are few, and although
%some HI deficient galaxies have been discovered in loose groups
%(Kilborn et al. 2005; Omar \& Dwarakanath 2005), their significance
%is not yet known.

This paper describes \HI\ observations of the NGC 3783 group, and in
particular, the discovery of several new group members, and a region
of HI emission that has no optical counterpart. This work is part of a
larger investigation into the properties of galaxies in groups, the
Group Evolution Multiwavelength Study (GEMS). GEMS is a large survey
of about 60 groups that were selected to have previous ROSAT PSPC
X-ray observations (see Forbes et al. 2006). Osmond \& Ponman (2004;
hereafter OP04) detail the selection criteria and group properties for
the 60 groups. We have made wide-field \HI\ observations of 16 GEMS
groups with the Parkes telescope. These observations will allow us to
characterise the \HI\ content of loose groups, and the relationship
between \HI\ content and X-ray emission. Kilborn et al. (2005, 2006)
describe the \HI\ sub-sample and characteristics.

The NGC 3783 group was first identified by Giuricin et al. (2000), who
found three galaxies including the well-studied Seyfert galaxy NGC
3783. The group lies at a distance of 36 Mpc (OP04); at this distance,
1\arcsec = 174.5 pc.  OP04 find extended X-ray emission surrounding
NGC 3783, out to a radius of 69 kpc (see Section~\ref{xray} for
details).

%Note the other group members according to giuricin et al are ESO 320-004 and ESO 378-012
%Is there any independent distance measurement to this group?

We briefly discuss our \HI\ observations and data reduction in
Section~2, and the results from the wide-field \HI\ imaging in
Section~3. High resolution observations of NGC 3783 itself, and a
nearby dwarf galaxy are presented in Section~4, and we
discuss the discovery of a region of intra-group \HI\ gas in
Section~5.  We finish with discussions in Section~6, and conclusions in
Section~7. Throughout we use a heliocentric velocity in the optical
convention.

\section{Observations and data reduction}
\label{sect:obs}

\begin{table} 
\centering
\caption{Parkes narrow-band datacube parameters for the NGC 3783
group. }
\label{tab:hicube} 
\begin{tabular}{lc}
\hline
Centre  [$\alpha(^{\rm h\,m}),\delta(\degr\,\arcmin)$(J2000)]  & 11:37,--37:53\\                
Gridded beam size [\arcmin]       & 15.5\\
Cube size [$\degr$]          & 5.5 $\times$ 5.5\\
%Total observing time [hr]  & 18.5\\
Velocity range [km s$^{-1}$]       & 1979 -- 3615\\
Channel width  [km s$^{-1}$]         & 1.65  \\
Velocity resolution [km s$^{-1}$]     & 2.6  \\
rms noise per channel [mJy beam$^{-1}$]   & 26  \\
\hline
\end{tabular}
\flushleft
\end{table}

\subsection{Parkes HI Observations}
\label{section:pks}

\begin{table*} 
\caption{Details of high resolution ATCA \HI\ data obtained for
galaxies in the NGC 3783 group. The columns are as follows: (1) GEMS
galaxy number; (2) New observation (O), or archive data used (A); (3)
Observation dates; (4) ATCA array used for the observations; (5) Total
integration time obtained for source in hours; (6) ATCA beam size in
arcsec$^{2}$; (7) Spectral resolution of the final datacube (\kms);
(8) rms noise level per channel of the final datacube (mJy
beam$^{-1}$). }
\label{tab:atca} 
\begin{tabular}{cclllcrr}
\hline
No. & O or A & Observation Dates &Array/s& Int. Time & Beam size & Spectral Res. & rms \\
    &        &                   &       &   (h)     & ($\arcsec \times \arcsec$) & (\kms) & (mJy beam${-1}$)\\
(1)   &   (2)    &   (3)              &(4)      &(5)  & (6) & (7)   & (8)        \\
\hline
1 & O & 2006 Jan 28 & 750D & 1.6 &102$\times$34 & 6.6 & 5.5 \\
2 & O &2004 March 03, 2006 Jan 25-26  & 750A, 750D & 20  & 88$\times$50 & 13.2 & 2.8\\
3 & O & 2006 Jan 27 & 750D & 1.35 &83$\times$47 & 6.6 & 4.0\\
4,5& A & 2001 July 09 & 375 & 1 & 58$\times$145 & 6.6 & 6.0\\
6 & O & 2004 Nov 12 & 750C & 4.5 &98$\times$55 & 3.3 & 5.0\\
7 & A & 1993 Oct 09-10, 1993 Sept 29-30, &1.5D, 750D, 1.5A& 18.1 &50$\times$50 &10.0 & 1.5\\
  &   & 1996 Oct 17-18                   &                &      & & \\

\hline\\

\end{tabular}
\end{table*}

\begin{figure*}
\begin{tabular}{ccc}
\mbox{GEMS\_N3783\_1} & \mbox{GEMS\_N3783\_2} & \mbox{GEMS\_N3783\_3}\\
 \mbox{\psfig{file=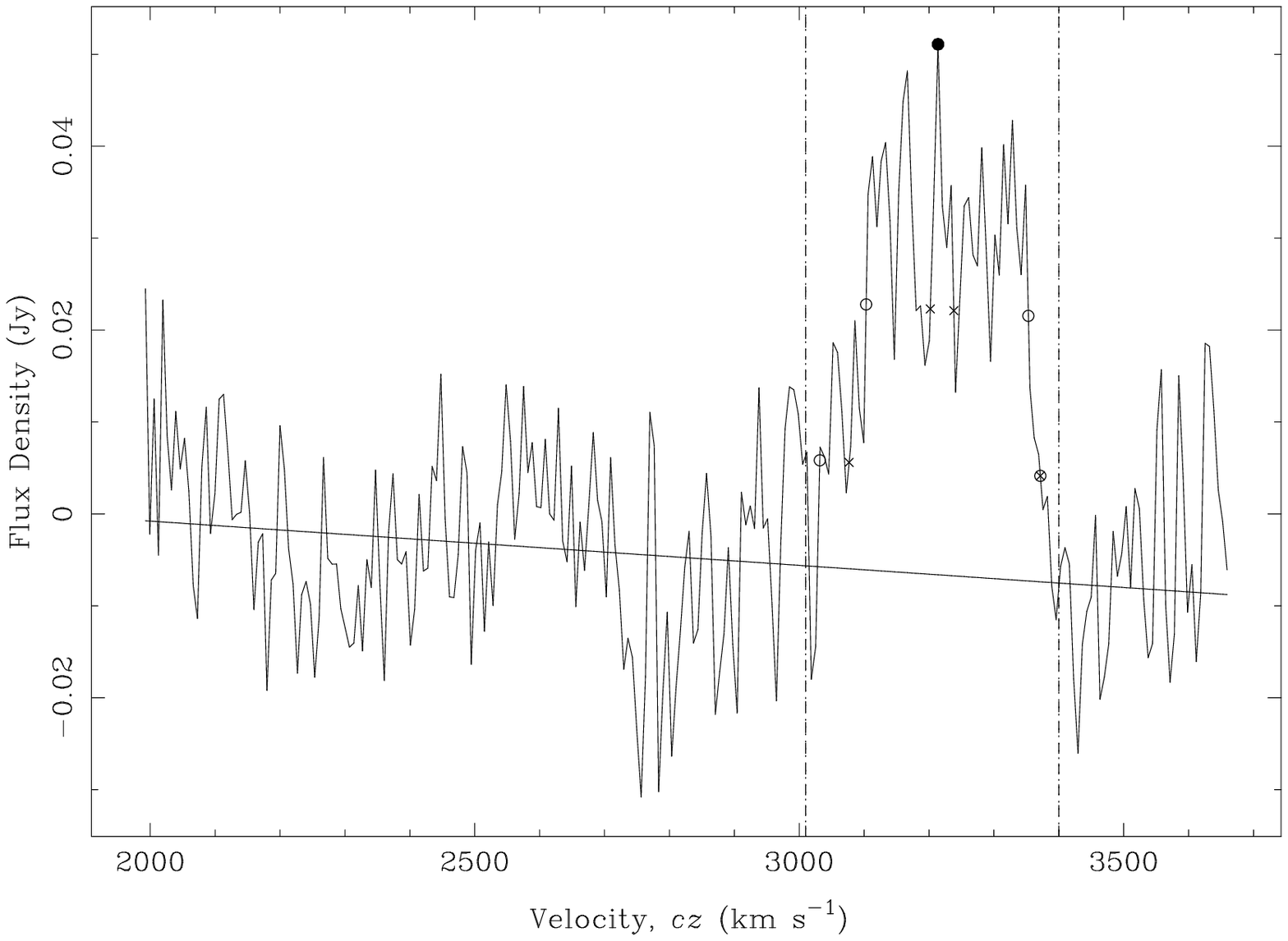,width=5.2cm}}&
 \mbox{\psfig{file=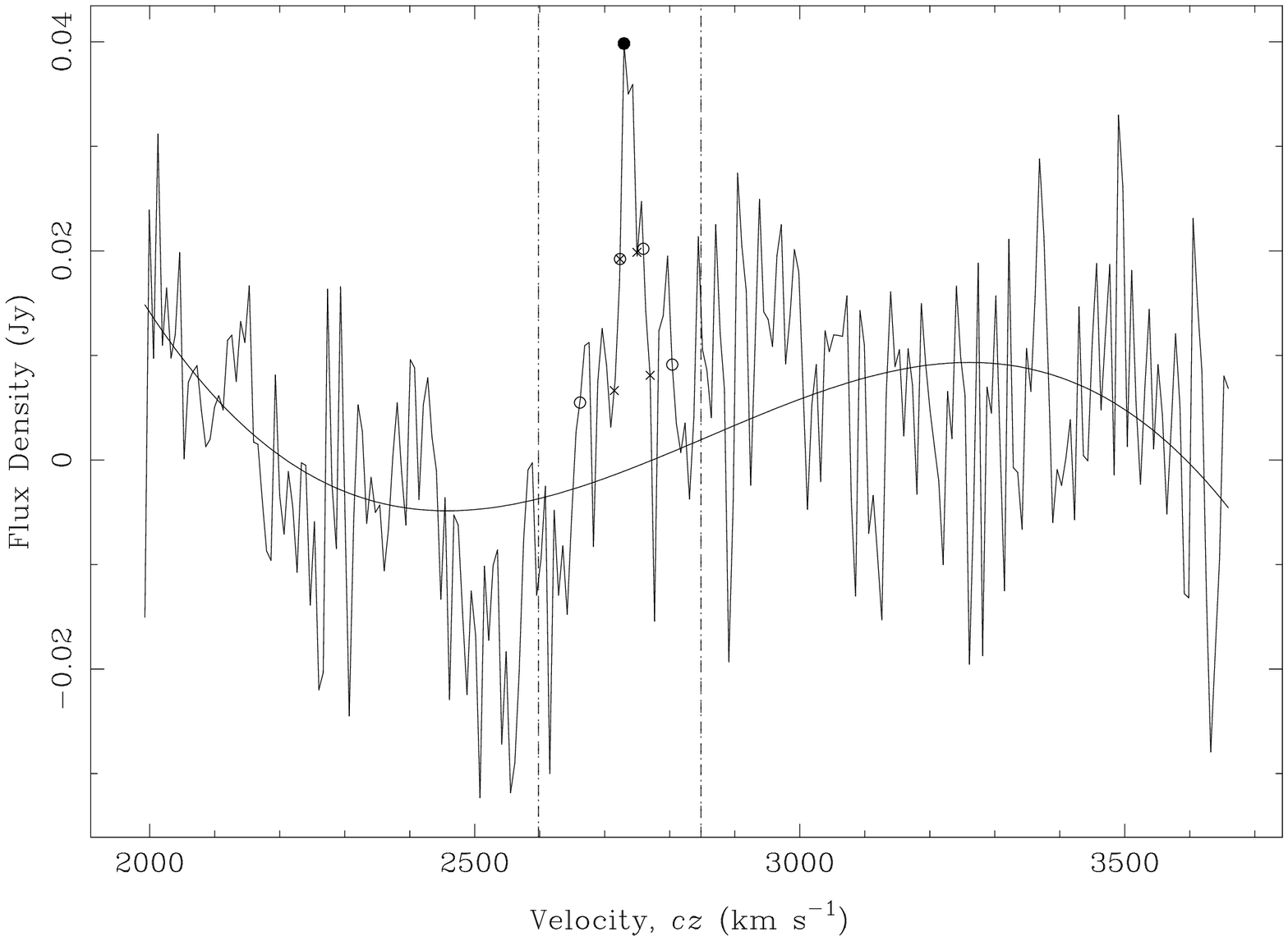,width=5.2cm}}&
 \mbox{\psfig{file=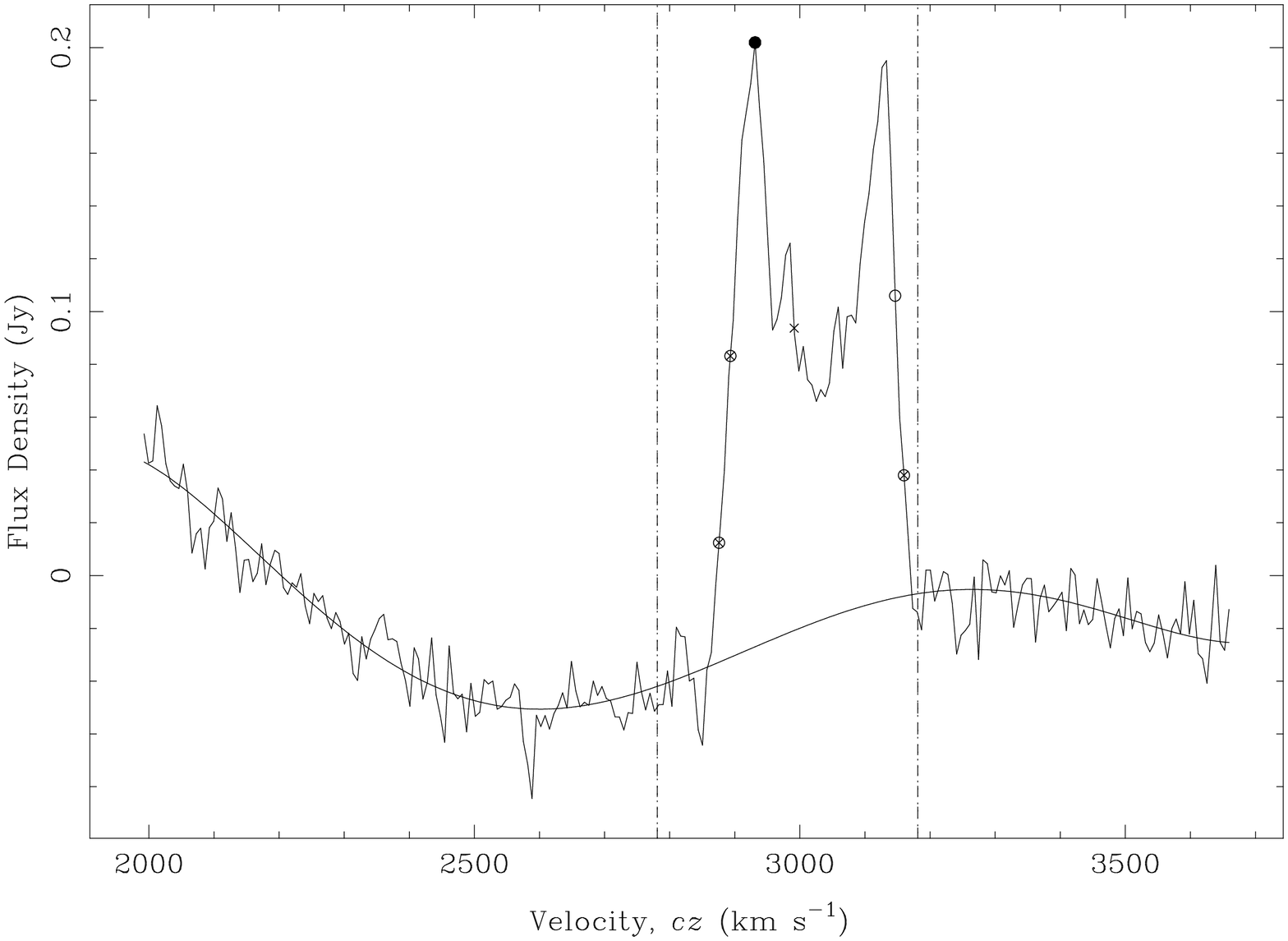,width=5.2cm}}\\ 
\mbox{} & \mbox{} & \mbox{}\\
\mbox{GEMS\_N3783\_4} & \mbox{GEMS\_N3783\_5} & \mbox{GEMS\_N3783\_6}\\
 \mbox{\psfig{file=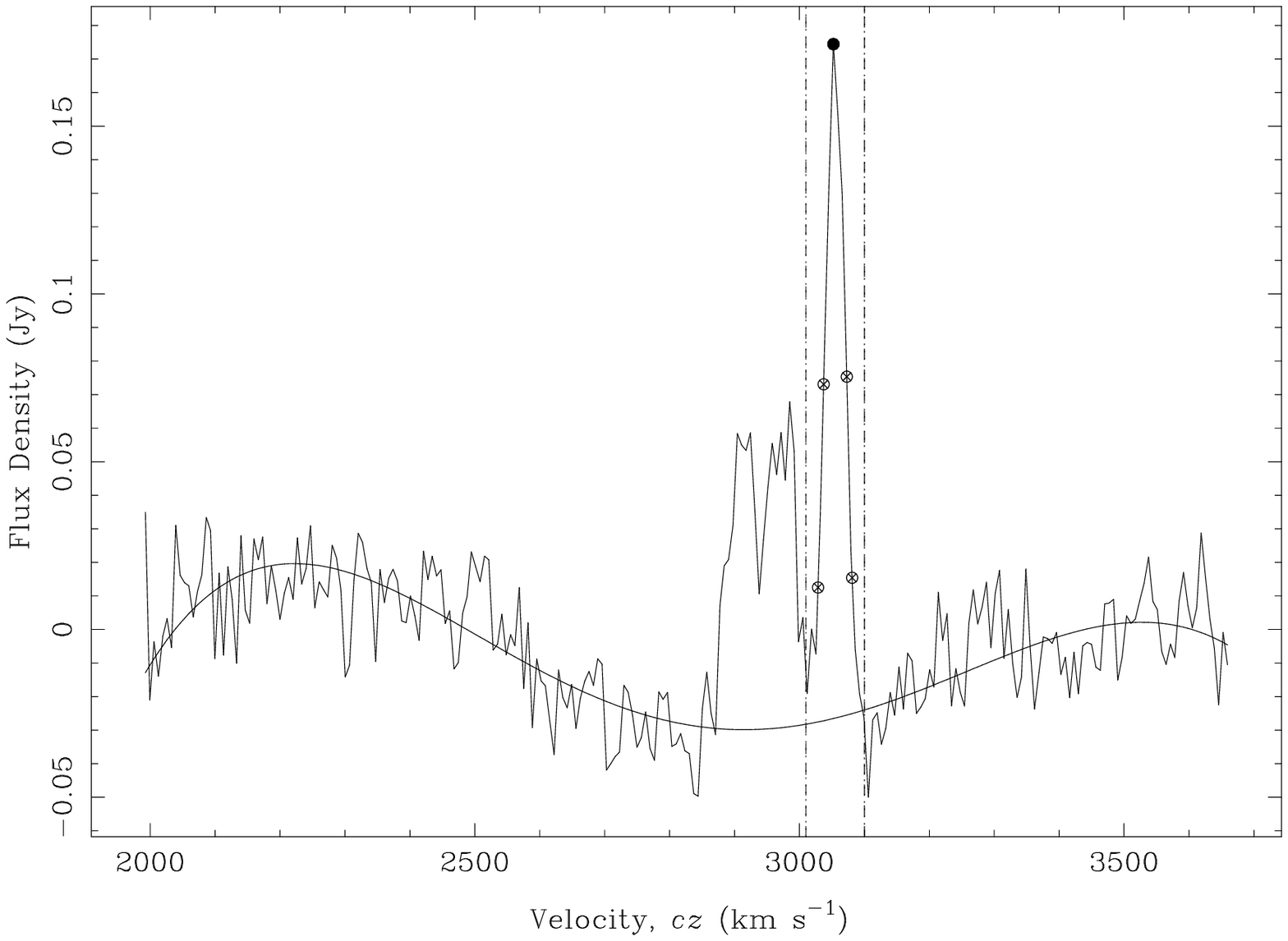,width=5.2cm}}&
 \mbox{\psfig{file=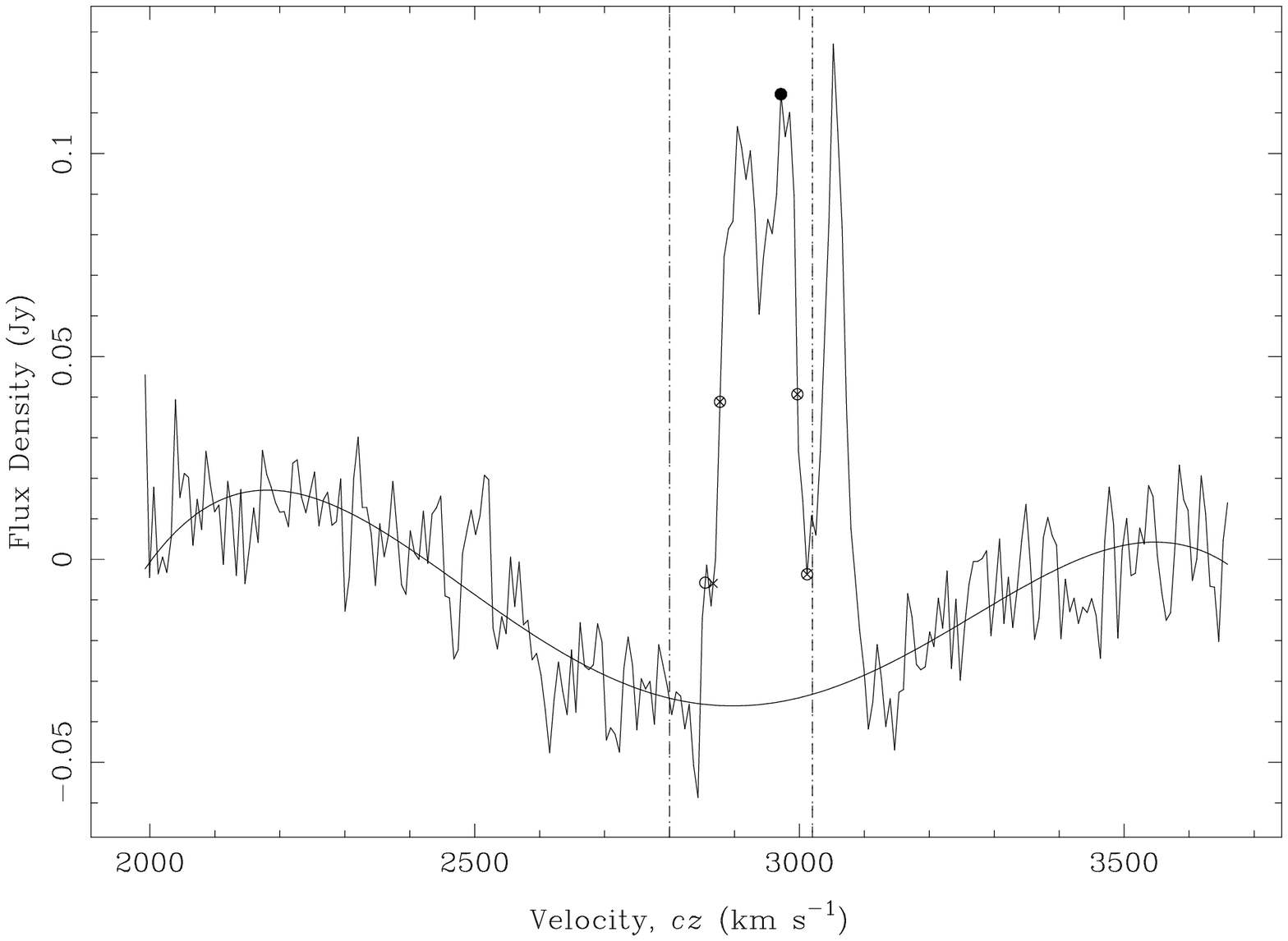,width=5.2cm}}& 
 \mbox{\psfig{file=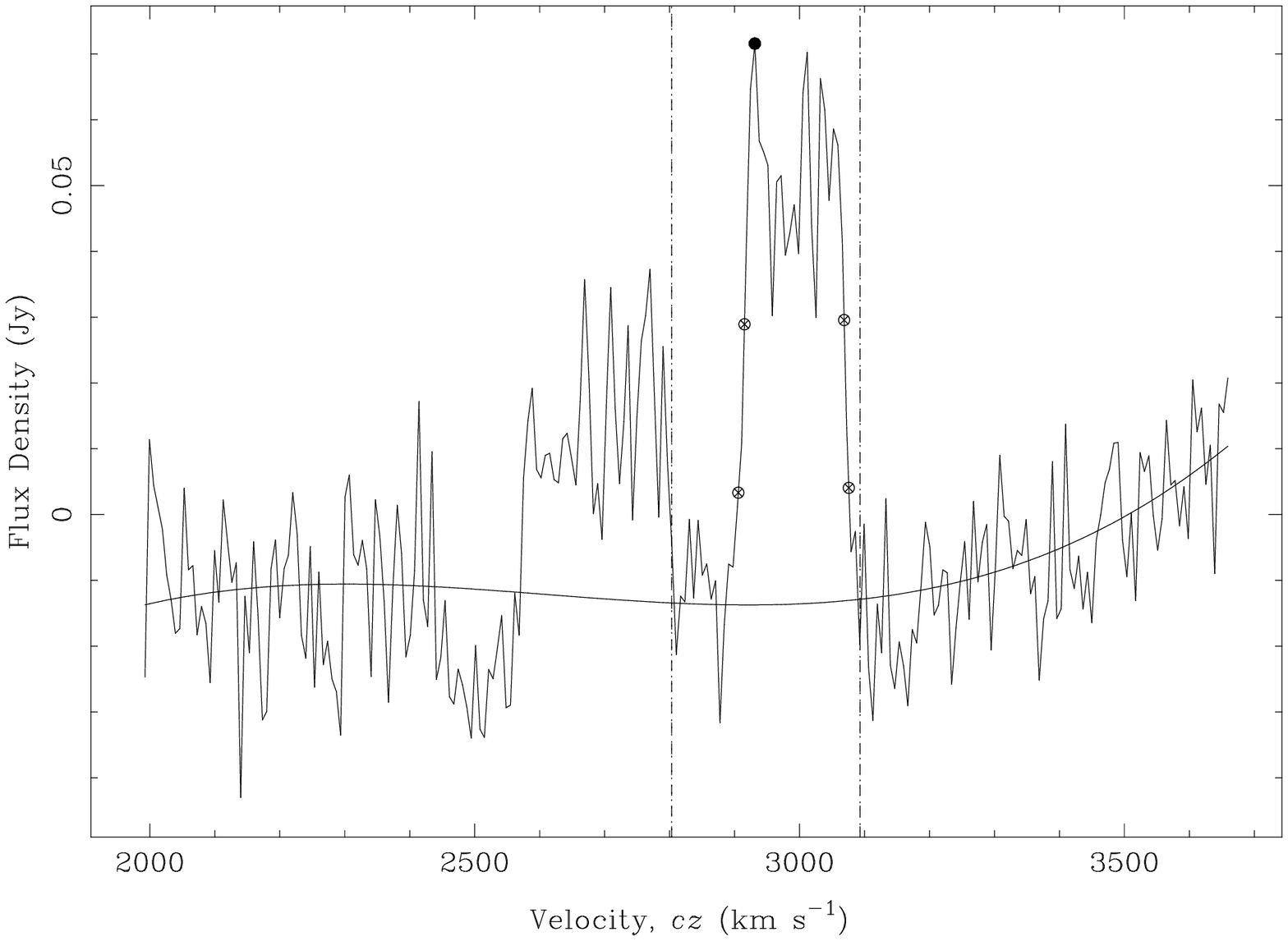,width=5.2cm}}\\
 \mbox{} & \mbox{} & \mbox{}\\
\mbox{GEMS\_N3783\_7} & \mbox{GEMS\_N3783\_8} & \mbox{GEMS\_N3783\_9}\\
 \mbox{\psfig{file=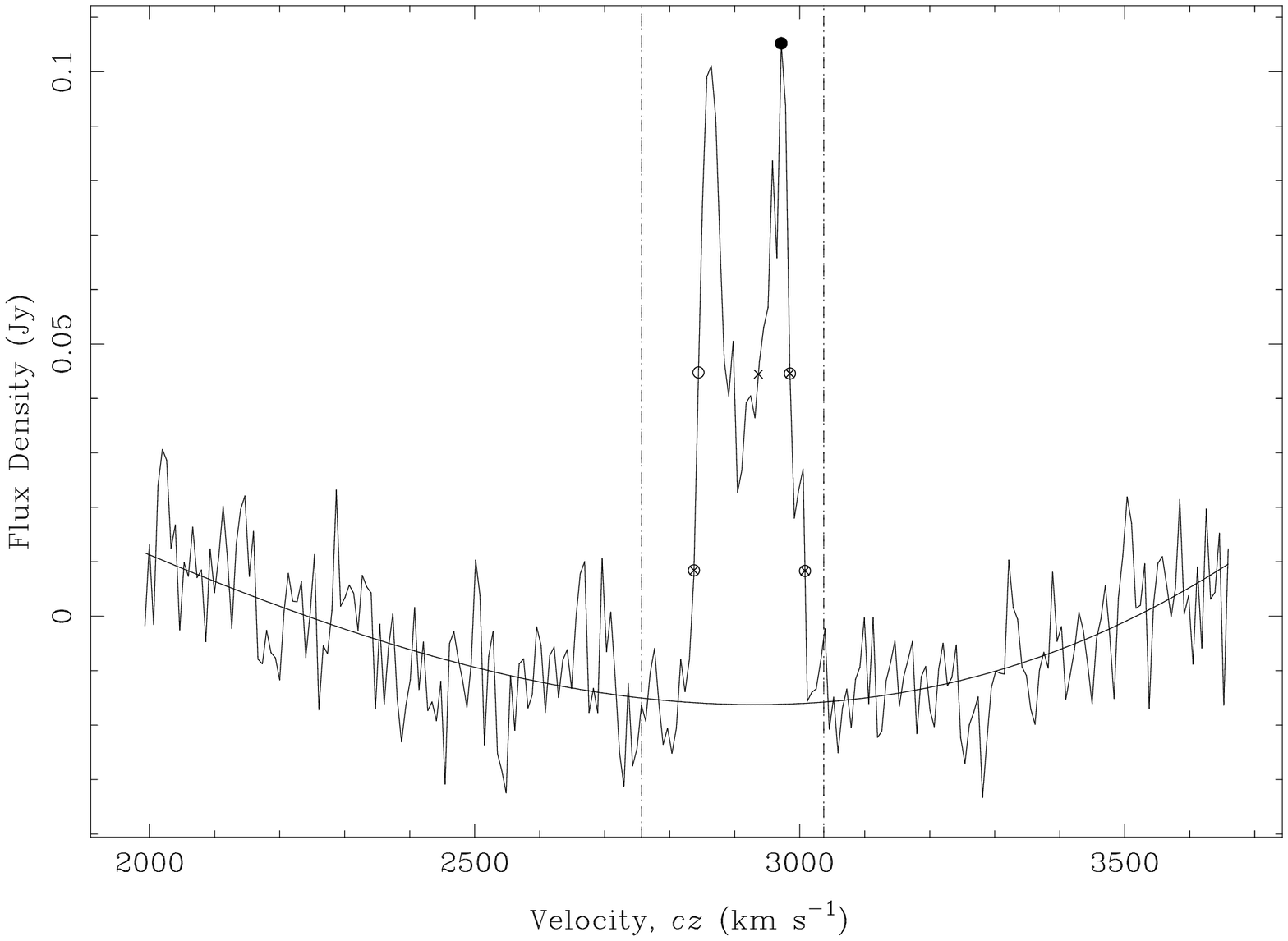,width=5.2cm}}&
 \mbox{\psfig{file=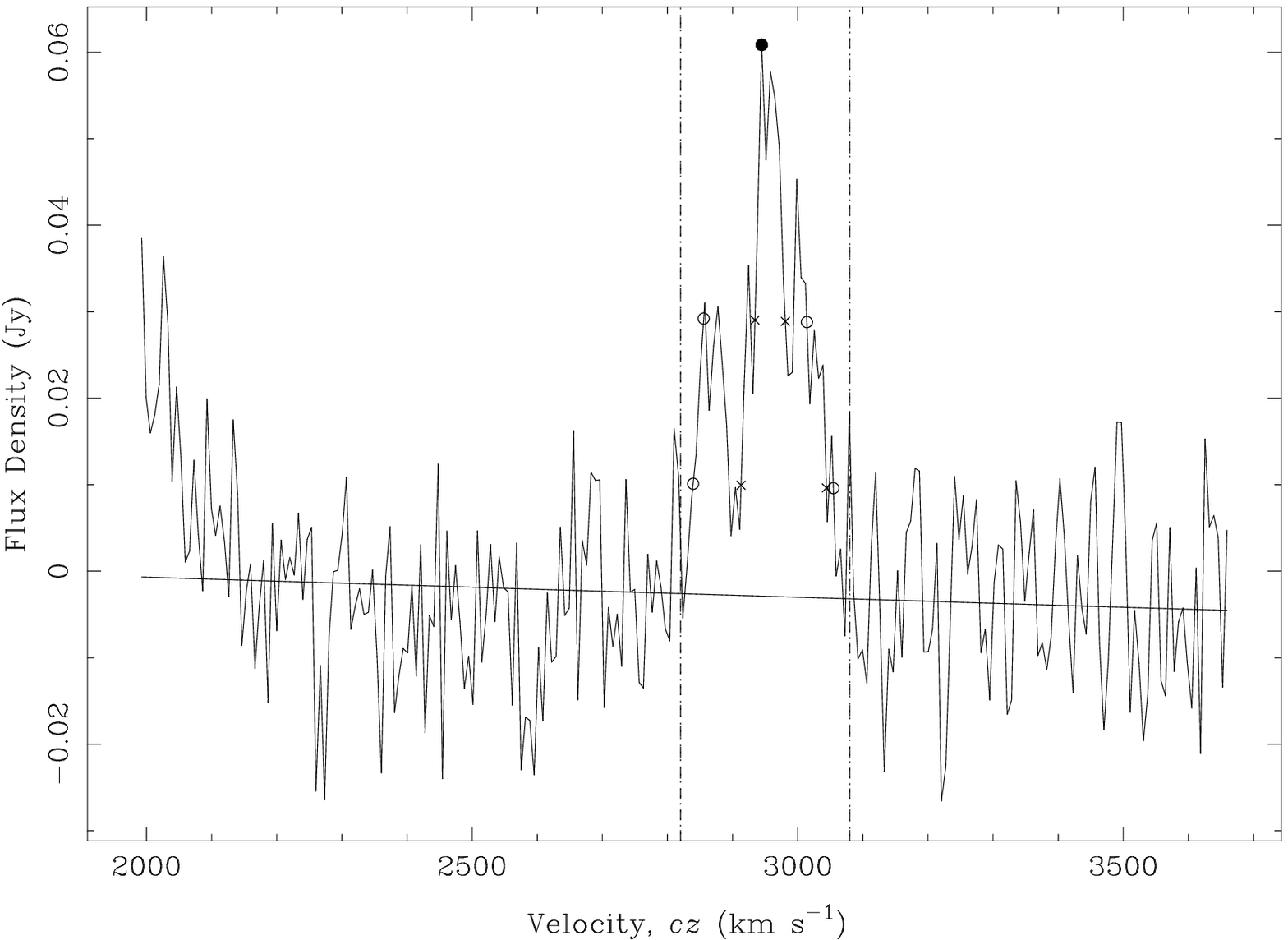,width=5.2cm}}& 
 \mbox{\psfig{file=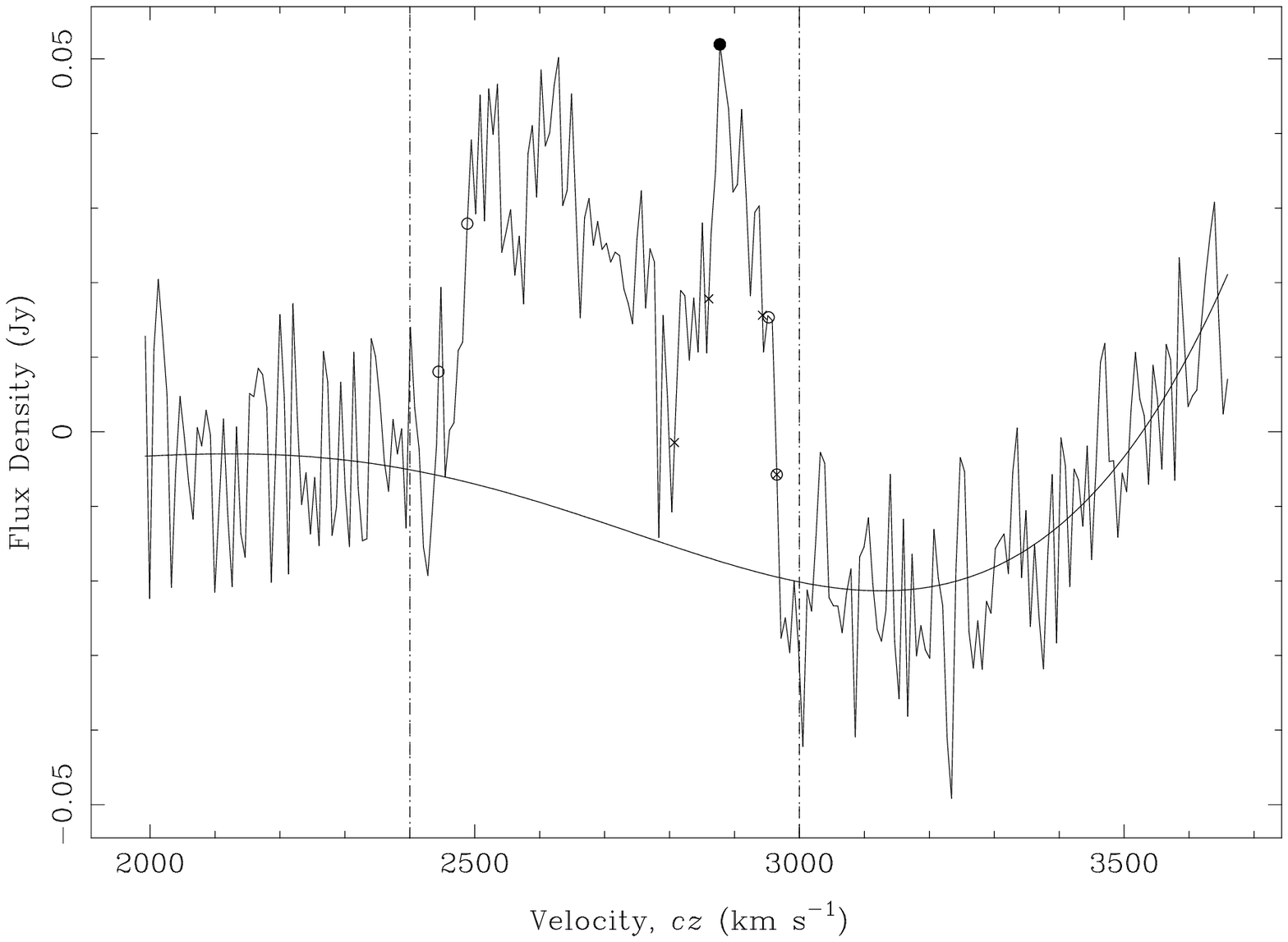,width=5.2cm}}\\
\mbox{} & \mbox{} & \mbox{}\\
\mbox{GEMS\_N3783\_10} & \mbox{GEMS\_N3783\_11} & \mbox{GEMS\_N3783\_12}\\
 \mbox{\psfig{file=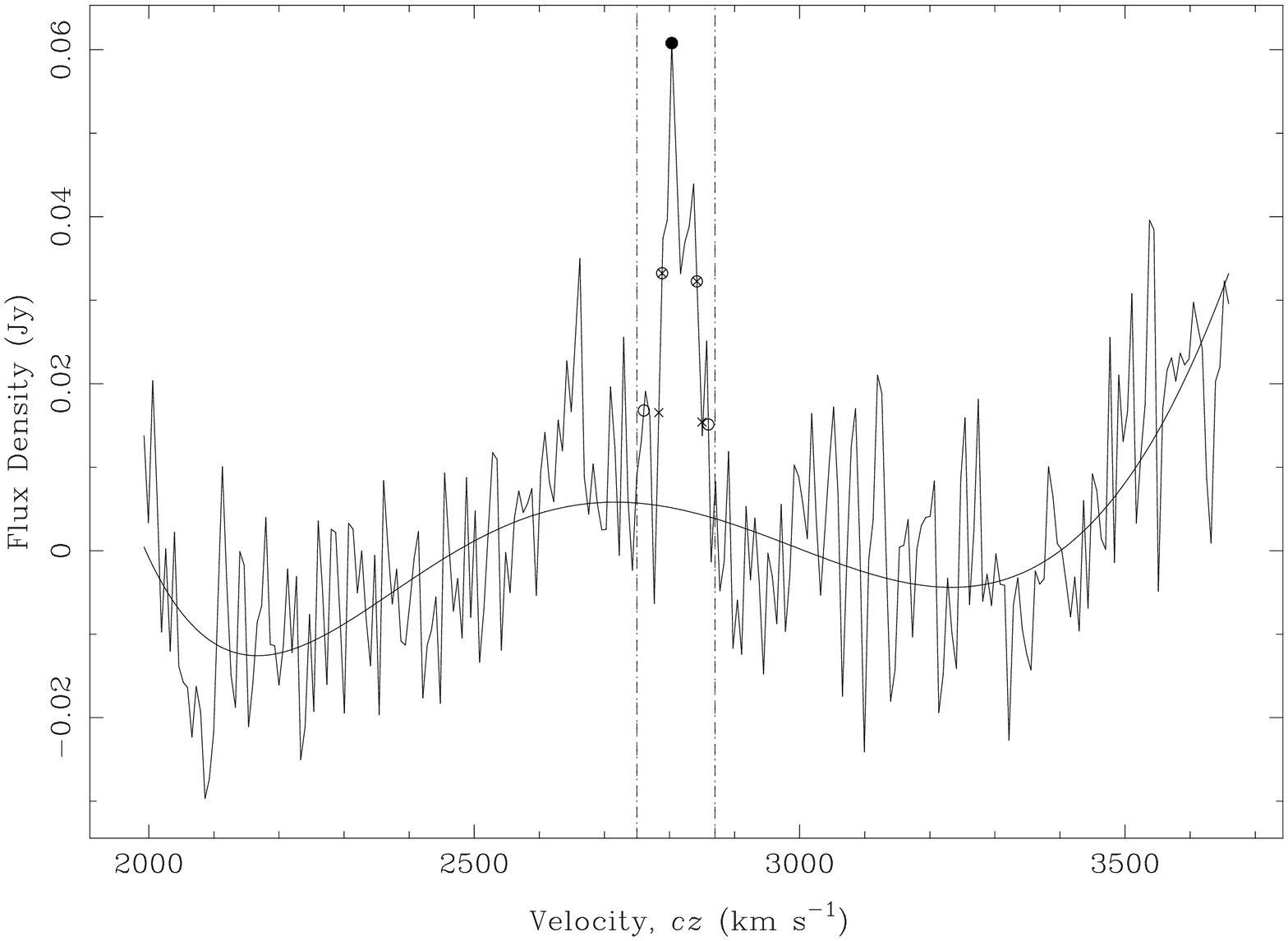,width=5.2cm}}&
 \mbox{\psfig{file=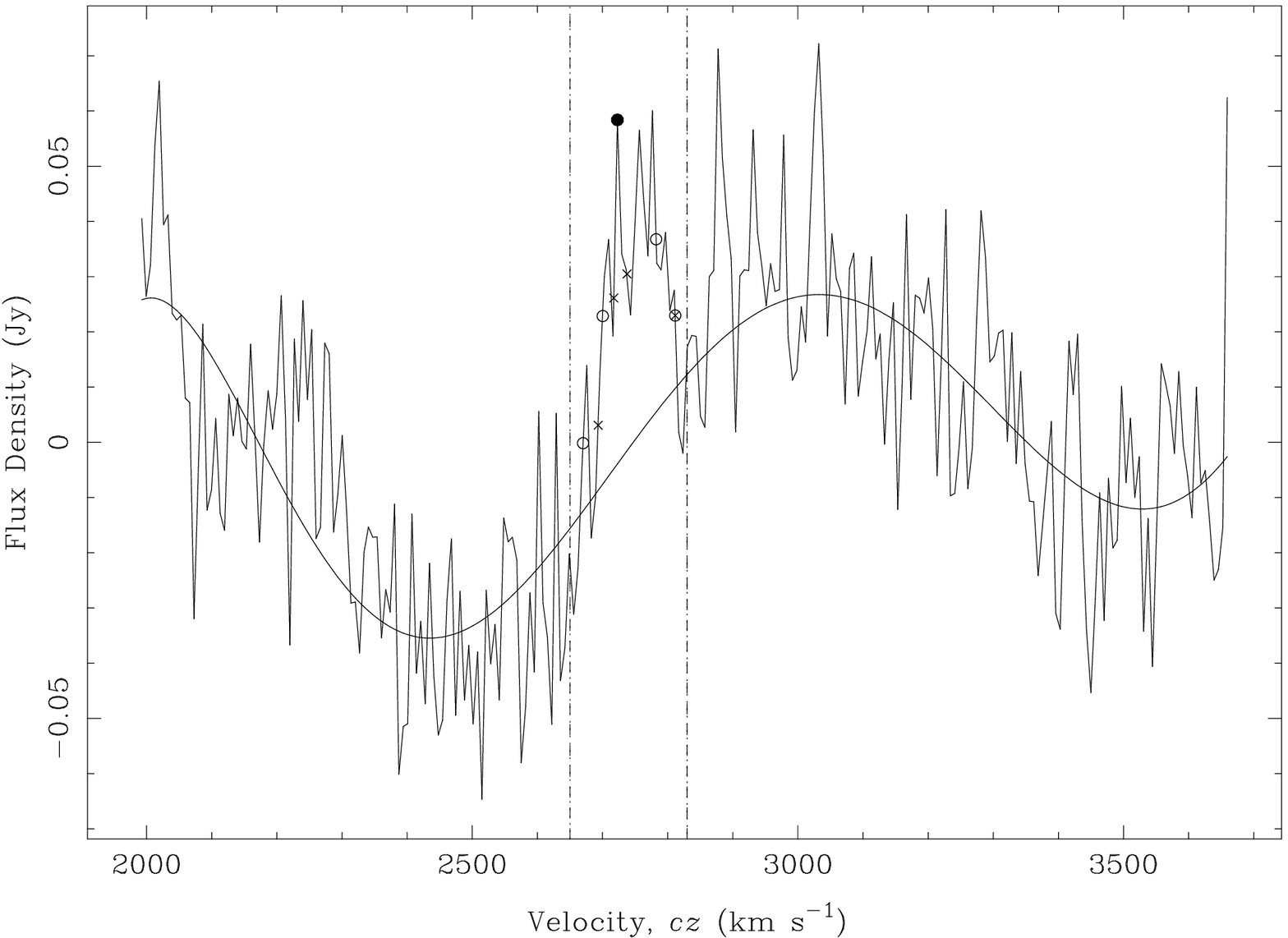,width=5.2cm}}& 
 \mbox{\psfig{file=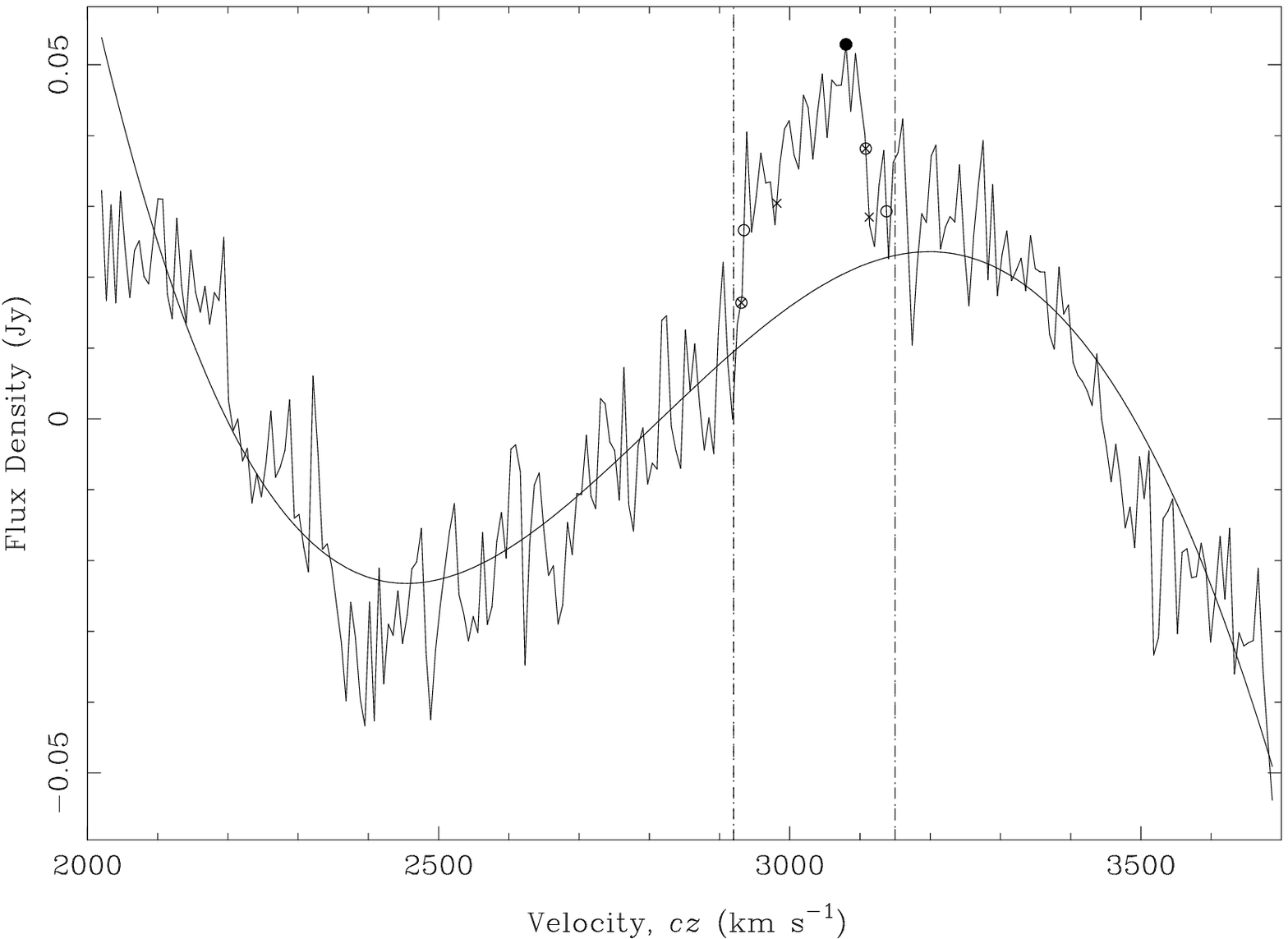,width=5.2cm}}\\
\end{tabular}
\caption{\HI\ spectra of the twelve galaxies detected in our
   NGC~3783 survey. The spectra were Hanning smoothed to a velocity
   resolution of 6.6 \kms. The fitted baseline is shown and the \HI\
   peak flux density is marked with a filled circle. The $w_{20}$ and
   $w_{50}$ velocity widths are shown by the open circles (outer fit),
   and crosses (inner fit). The velocity region between the vertical
   lines in the spectra was disregarded in the baseline fit.}
\label{fig:hispectra}
\end{figure*}

Wide-field \HI\ observations for the NGC 3783 group were conducted as
part of the GEMS \HI\ study (Kilborn et al. 2005, 2006). The
observations were made using the inner seven beams of the 20-cm
multibeam system on the Parkes 64-m telescope in NSW, Australia. The
area observed covers $\sim 5.5^{\rm o} \times 5.5^{\rm o}$ centred on
the NGC 3783 galaxy group. The velocity range of the data is from 1979
to 3615 km s$^{-1}$.  The observational strategy and subsequent data
reduction are described in detail in Kilborn et al. (2005,
2006). Approximately 25 hours of data were obtained for this
region. We used the aips++ routines {\sc livedata} and {\sc gridzilla}
to reduce the data, in the same way that is explained in detail in
Kilborn et al. (2005).

 For this particular dataset, approximately 30\% of the data was
affected by solar interference, which produces a `ripple' effect in
the reduced datacube. While the \HI\ detections in the cube are
visible, the cube contains a number of artifacts and the data quality
is variable both spatially and in velocity. Thus, a subsequent
datacube was made from the remaining 70\% of ripple-free scans
only. Although the rms noise for this datacube is slightly higher, the
artifacts that had appeared in the original datacube are absent, and
the rms is constant throughout the new datacube.  The new,
ripple-free, \HI\ datacube has an rms noise of 26 mJy beam$^{-1}$ per
channel, and the parameters for the datacube are given in
Table~\ref{tab:hicube}. We smoothed this datacube to a velocity
resolution of 13.2 \kms, and the rms noise in the smoothed cube is
11.8 mJy beam$^{-1}$ per channel. For the distance of this group, this
gives a 3$\sigma$ \HI\ mass limit of 3.8$\times 10^8$\Msun\ (assuming
a Gaussian profile, and an \HI\ velocity width of 50 \kms). The
smoothed \HI\ datacube was searched for \HI\ detections visually,
using the {\sc karma} visualisation package (see Kilborn et al. 2005,
2006, for cube searching and source parameterisation details).

%We made confirmation observations of all detected sources in the
%Parkes \HI\ datacube. These observations were carried out at the
%Parkes radiotelescope on 2005 October 26-31. The observations were
%made in 'MX' mode, where the telescope is pointed at the source, and
%the inner seven beam are sequentially moved on and off the source,
%enabling the source and reference beams to be observed at the same
%time. The total integration time for each observation was 20 minutes.
%The observations used an 8 MHz bandpass, with 2048 channels. The data
%were bandpass removed in livedata, and gridded using gridzilla. A
%calibration observation was made for each observation by observing the
%standard flux calibrator 1934-638. 

\subsection{ATCA Observations}
\label{n3783obs}

We obtained higher spatial resolution observations from the Australia
Telescope Compact Array (ATCA) both through dedicated observing
programs, and from the ATCA archive.  Standard {\sc miriad} routines
were applied to reduce the data, and `natural' weighting was used in
each case. We used the primary flux calibrator 1934-638 for all the
observations. The full-width half-power primary beam of the ATCA is
$\sim 33\arcmin$ at 1.4 GHz. The observational details are summarised
in Table~\ref{tab:atca}.

\begin{table*} 
\caption{\HI\ detections from our NGC 3783 Parkes datacube.}
\label{tab:hi} 
\begin{tabular}{lcrrcll}
\hline
No.  & $\alpha,\delta$(J2000)&$v_{sys}$& \MHI &Separation &Optical ID/s & Velocity\\
         & [$^{\rm h\,m\,s}$], [\degr\,\arcmin\,\arcsec]
          & [\kkms] &  [$10^8 $\Msun] &[arcmin] & & [\kms]\\
(1) & (2) & (3) & (4) & (5) &(6)& (7)\\
\hline
1 &11:34:18 -37:09:39 & 3202  $\pm$ 7 & 34.0 $\pm$3.3 & 6.1      & ESO 378- G 011          & 3245      \\
2 &11:31:32 -36:18:53 & 2733  $\pm$10 &  3.8 $\pm$1.3 & $\cdots$ & $\cdots$                & $\cdots$  \\
3 &11:28:06 -36:32:45 & 3018  $\pm$ 2 &116.0 $\pm$5.4 & 0.4      & ESO 378- G 003          & 3022      \\
4 &11:49:40 -37:30:50 & 3055  $\pm$ 1 & 21.0 $\pm$2.3 & 1.7      & AM1147-371              & 2964      \\
5 &11:49:01 -37:29:47 & 2933  $\pm$ 2 & 51.1 $\pm$3.4 & 1.3      & ESO 378- G 023          & 2932      \\
  &                   &               &               & 2.0      & NGC 3903                & 2983      \\
6 &11:49:33 -38:50:40 & 2991  $\pm$2  &32.5  $\pm$2.9 & 1.8      & ESO 320- G 024          & 3037      \\
  &                   &               &               & 4.9      & ESO 320- G 026          & 2716      \\
  &                   &               &               & 5.5      & 6dF J1149529-385431     & 2708      \\
7 &11:38:51 -37:47:38 & 2923  $\pm$2  &38.0  $\pm$2.9 & 3.9      & NGC 3783                & 2817      \\
8 &11:37:54 -37:56:04 & 2947  $\pm$5  &21.0  $\pm$2.5 & $\cdots$ & $\cdots$                & $\cdots$  \\
9 &11:35:43 -38:02:08 & 2705  $\pm$5  &59.5  $\pm$4.1 & 1.0      & AM 1133-374             & 2742      \\     
  &                   &               &               & 3.0      & NGC 3749                & 2742      \\
  &                   &               &               & 5.2      & NGC 3742                & 2715      \\
  &                   &               &               & 8.6      & 2MASX J11351493-3755309 & 2823      \\
10&11:26:06 -37:51:26 & 2810  $\pm$5  & 7.2  $\pm$1.5 & 4.0      & ESO 319- G 020          & $\cdots$  \\
11&11:21:57 -37:46:45 & 2740  $\pm$5  &12.6  $\pm$2.0 & 7.2      & ESO 319- G 015          & 2737      \\
12&11:29:43 -37:16:59 & 3034  $\pm$7  &13.7  $\pm$2.7 & 4.3      & ESO 378- G 007          & 3041      \\

\hline
\end{tabular}
\flushleft The columns are (1) GEMS galaxy number, (2) fitted \HI\
centre position, (3) \HI\ systemic velocity in the optical convention,
(4) \HI\ mass for the detection using group distance of 36 Mpc. (5)
Distance of optical counterpart from centre of the HI emission.  (6)
Optical counterpart/s for the \HI\ detection (7) Velocity of the
optical counterpart, from 6dFGS DR2, apart from AM1147-371 where the
velocity is a previous \HI\ measurement from Matthews, Gallagher \&
Littleton (1995).  The errors are derived following Koribalski et
al. (2004). Cols (2)-(4) are \HI\ properties as derived from the
Parkes data.\\

\end{table*}

\section{HI content and Characteristics of the NGC 3783 group}
\subsection{\HI\ Characteristics of the group}

Figure~\ref{fig:hi_mom0} shows a velocity integrated \HI\ map from the
Parkes data for the NGC 3783 group. The \HI\ observations reveal a
very loose grouping, that is not symmetrically distributed around the
galaxy NGC 3783.  We detect twelve sources in our Parkes \HI\
datacube, including the galaxy NGC 3783 itself.  Table~\ref{tab:hi}
lists the \HI\ position, systemic velocity and \HI\ mass (assuming a
distance of 36 Mpc) for each of the NGC 3783 group members as measured
in the Parkes HI datacube, as well as their optical counterparts and
6dFGS DR2 velocity (see below). Their \HI\ spectra are shown in
Figure~\ref{fig:hispectra}, and additional \HI\ properties will be
presented in Kilborn et al. (2006).

We used the NASA Extragalactic Database (NED) and the 6dF Galaxy
Survey Data Release 2 (6dFGS DR2; Jones et al. 2005) to find
previously catalogued galaxies in the region of the \HI\
datacube. Those galaxies in the region with a known redshift between
2500--3500 \kms\ are indicated with a cross on
Figure~\ref{fig:hi_mom0}. We searched a region of 10 arcminutes around
the position of each \HI\ detection to find the corresponding optical
counterparts. These are listed in Table~\ref{tab:hi}.

Two of the \HI\ detections (GEMS\_N3783\_2, and GEMS\_N3783\_8) were
not previously catalogued in any optical catalogues (see
Section~\ref{newgal} and Section~\ref{cloud} for details). One \HI\
detection, GEMS\_N3783\_10, provides the first velocity for ESO 319- G
020, placing it at the group velocity. Details of the new group members
are summarised in Table~\ref{tab:newgals}.

There were four \HI\ detections that have several optically catalogued
sources (with velocities near that of the \HI\ source) within our 10
arcminute search region. Those \HI\ detections are GEMS\_N3783\_6,
GEMS\_N3783\_4, GEMS\_N3783\_5 and GEMS\_N3783\_9. As described in
Section~\ref{n3783obs}, we observed GEMS\_N3783\_6 in higher spatial
resolution at the ATCA. We found the \HI\ was associated with at least
two of the galaxies close to the central position of the Parkes \HI\
detection (ESO 320-G 023 and ESO 320-G 026), and a tentative detection
was made of the other galaxy in the region, 6dF
J1149529-385431. Archive ATCA data for GEMS\_N3783\_4 and
GEMS\_N3783\_5 was available. These two \HI\ detections lie very close
to one another in projection (separation of $\sim$ 2\arcmin), but are
separated in velocity in the Parkes data by 114 \kms. The ATCA
observations confirm that GEMS\_N3783\_4 is associated with
AM1147-371. However, the beam size of the ATCA observations was such
that it was not possible to confirm whether GEMS\_N3783\_5 is
associated with one, or both of ESO 378-G 023 and NGC 3903, so we list
both in Table~\ref{tab:hi}. We do not have any high resolution imaging
of GEMS\_N3783\_9, so we list all 4 nearby optical candidates in
Table~\ref{tab:hi}.

The HIPASS catalogue, HICAT, contains eight of the twelve galaxies we
detected in our \HI\ datacube (Meyer et al. 2004). Three of the
detections that are not in HICAT have small \HI\ fluxes, and the
fourth, GEMS\_N3783\_5, appears to have been missed in HICAT due to
confusion with the nearby GEMS\_N3783\_4. The HIPASS rms in the region
of the NGC 3783 group is $\sim 15-16$ mJy beam$^{-1}$ per channel,
which higher than our GEMS data that has an rms noise of 11.8 mJy
beam$^{-1}$ per channel (smoothed to the same resolution as HIPASS).

We compare the \HI\ fluxes between HICAT and GEMS and find there is
excellent agreement for six of the eight galaxies that the two surveys
have in common. There is a disagreement in \HI\ flux for two of the
galaxies that are in both HICAT and GEMS.  The GEMS \HI\ flux
measurement for GEMS\_N3783\_6 is $\sim $45 per cent lower than the
HICAT flux measurement (HIPASSJ1149-38a). In this case, the HICAT
spectrum has an extremely uneven baseline and thus the GEMS flux is
more reliable.  In the case of GEMS\_N3783\_9, the GEMS flux is 30\%
higher than the HICAT flux (HIPASSJ1135-38). In this case, both GEMS
and HICAT \HI\ spectrum have a good baseline so it is unclear what
causes the discrepancy in the flux. An independent measurement will be
needed to confirm the flux for this source. As our rms noise is lower,
we adopt the GEMS derived flux for this source.

%In two cases where there is no HICAT
%flux measurement, we have an independent ATCA measurement
%(GEMS\_N3783\_2 and GEMS\_N3783\_8). The \HI\ flux from Parkes is
%twice that of the ATCA for GEMS\_N3783\_2, and the GEMS flux is 60\%
%higher than the ATCA measurement for GEMS\_N3783\_8, however it is
%likely that a significant amount of \HI\ flux was resolved out with
%the higher resolution, ATCA observations.  A further comparison with
%GEMS and HIPASS values can be found in Kilborn et al. (2006).

\subsection{X-ray characteristics of the group}
\label{xray}

\begin{figure} 
\begin{tabular}{c}
\mbox{\psfig{file=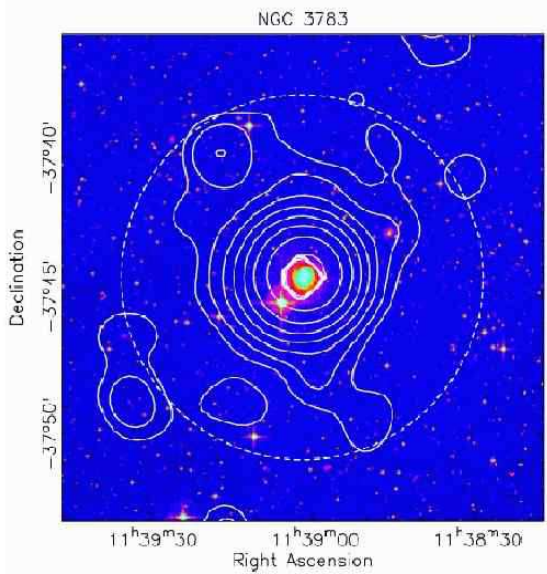,height=8cm}}
\end{tabular}
\caption{X-ray contours overlaid on the optical image of NGC 3783. The
  dashed circle indicates the detectable extent of the group X-ray
  emission (Forbes et al. 2006).}
\label{fig:xray}
\end{figure}

X-ray observations of the NGC 3783 group were obtained from the ROSAT
PSPC archive, and the data reduction is described in OP04. X-ray
images for each galaxy group in the GEMS survey are available in
Forbes et al. (2006). The resolution of the X-ray images are
$30\arcsec$. The X-ray emission is centred on the galaxy NGC 3783, and
OP04 find NGC 3783 itself to have an extended X-ray halo, consistent
with intra-group X-ray emission (see Figure~\ref{fig:xray}). However,
they were unable to fit a 2-component model to the X-ray distribution,
and thus were unable to distinguish the {\it galaxy} emission from the
{\it group} X-ray emission. The extent of the X-ray emission in NGC
3783 is 69 kpc, which is low compared to other loose groups that
typically have group X-ray halos greater than 100 kpc in size
(Mulchaey \& Zabludoff 1998). The X-ray luminosity for NGC 3783 is low
at log $L_X$ = 40.76$\pm$0.11 ergs s$^{-1}$.

\subsection{Dynamical and optical properties of the NGC 3783 group}

\begin{figure*} 
\begin{tabular}{c}
\mbox{\psfig{file=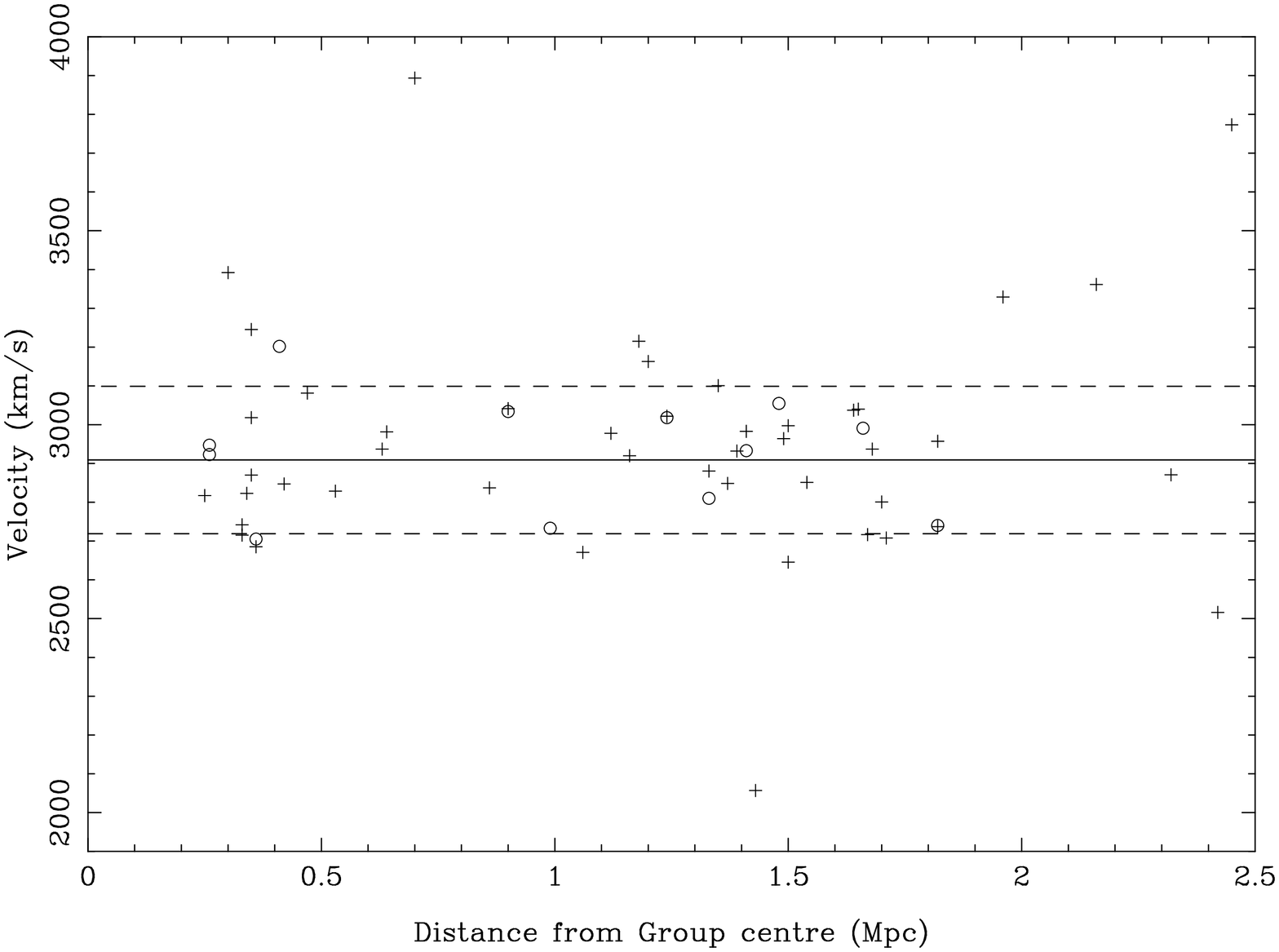,height=7cm,angle=-90}}
\mbox{\psfig{file=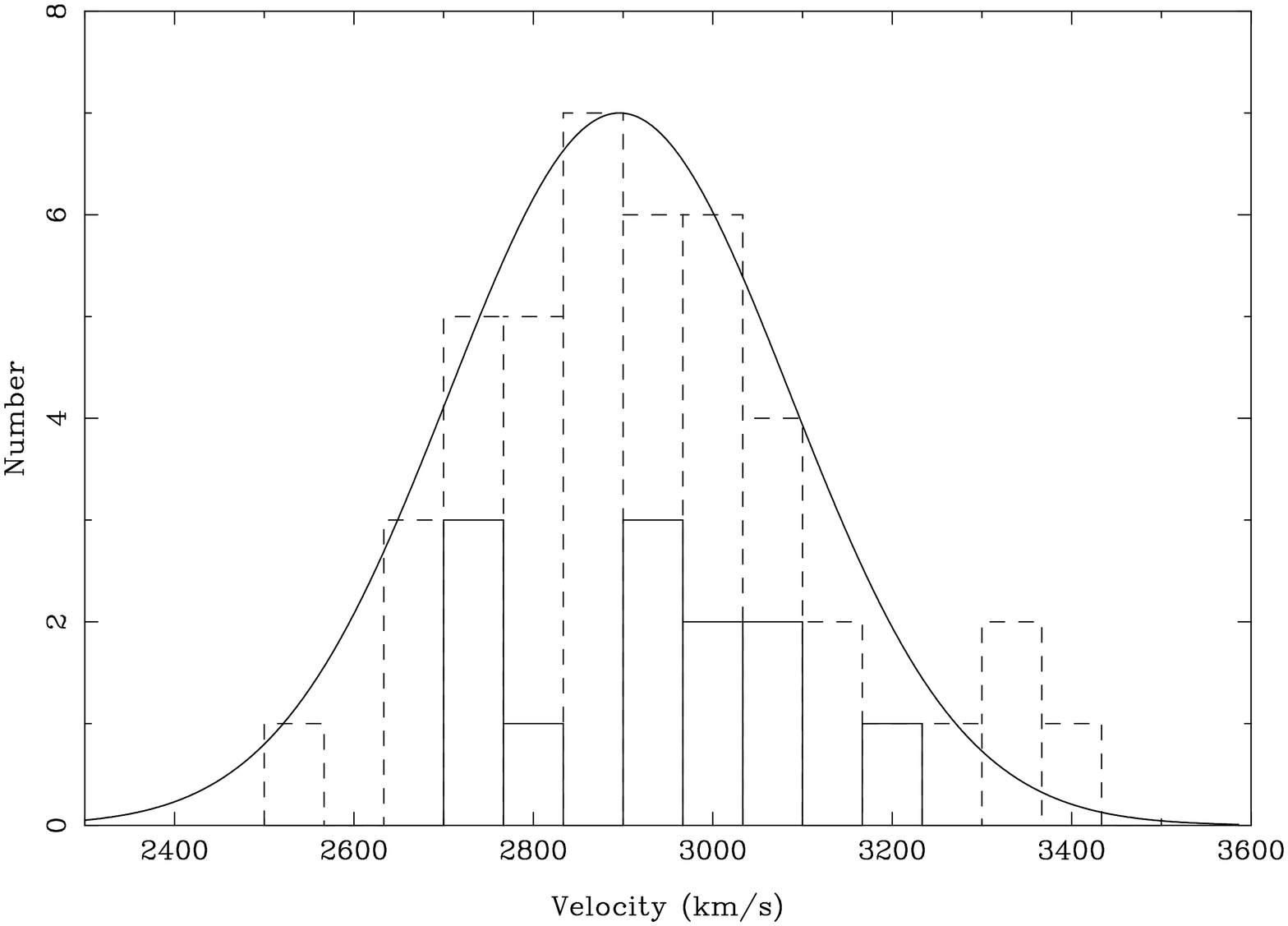,height=7cm,angle=-90}}
\end{tabular}
\caption{Left: Velocity-distance plot for the NGC 3783 group, using the
luminosity-weighted centre. The crosses mark the NED and 6dFGS DR2
sources, and the open circles are the \HI\ detections.The solid line
is the mean velocity of the group, 2903 \kms, and the dashed lines are
the velocity dispersion, 190 \kms. Right: Velocity histogram of the
NGC~3783 group, with Gaussian of width 190 \kms, and centre 2903 \kms\
overlaid. The dashed line shows the velocity distribution for all galaxies
in the region, and the solid histogram shows the \HI\ detected
sources.  }
\label{fig:veldist}
\end{figure*}

Using the NED and 6dFGS DR2 databases, we find 44 galaxies within the
 region of our \HI\ datacube, and between velocities 2500--3500
 \kms. Of these, 31 ($\sim 70$\%) have their only redshift from the
 6dFGS DR2. We use all previously catalogued galaxies in the region,
 along with the new HI detections, to investigate the characteristics
 of this group. The velocities used in the calculations are 6dFGS DR2
 velocity where available (79\%), then velocities from NED where
 available (15\%) and finally \HI\ velocities (6\%). Parameters for
 the NGC 3783 group are shown in Table~\ref{tab:higroup}.

To determine the centre of the NGC 3783 group, we use a
luminosity-weighted mean, based on the $K-$band magnitudes of the
galaxies where available from 2MASS. Those galaxies without a known
$K-$band magnitude are given a magnitude of 13.5 (as in Brough et
al. 2006). All galaxies within the extent of the \HI\ cube, and with
velocities $2500 < v_{sys} < 3500$ \kms\ are included in the
calculation. The luminosity weighted centre is $\alpha,\delta$(J2000)
= 11:37:12,$-$37:30:57.6, and is marked on
Figure~\ref{fig:hi_mom0}. This centre is closest to the galaxy NGC
3783 itself, at a distance of 267 kpc away. As NGC 3783 is the largest
galaxy in the group, and with an extended X-ray halo it is expected to
lie closest to the group dynamical centre.

Figure~\ref{fig:veldist} shows the velocity-distance distribution for
the NGC 3783 group, with the mean velocity and velocity dispersion for
the group overlaid. The mean velocity and velocity dispersion were
calculated following Beers et al. (1990).  We calculate a mean
velocity of 2903$\pm26$ \kms\ and dispersion of 190$\pm24$ \kms\ for
the group. The velocity of the galaxy NGC 3783 is very close to the
mean velocity of the group, at 2916 \kms\ (from the \HI\
data). Comparing these values to those in the literature, in the
original determination of the group, Giuricin et al. (2000) found a
median group velocity of 2854 \kms, which is slightly lower than our
value. OP04 search for group members for NGC 3783 using NED galaxies
within a radius of 0.25 Mpc (the radius at which the density of the
group fall to 500 times the critical density of the Universe,
calculated from the X-ray temperature of the group), and $\pm 3\sigma$
of the velocity dispersion of the group. However, under these strict
criteria, they find no other catalogued galaxies lie in the group
region. They quote the group velocity as 2917 \kms, which is the
optical velocity of the galaxy NGC 3783 itself. Brough et al. (2006)
use a friends-of-friends algorithm to determine the members and
characteristics of the NGC 3783 group, and find the group has nine
members, a mean group velocity of 2826$\pm14$ \kms, and velocity dispersion
of 118$\pm37$ \kms.

The velocity distribution of the galaxies is also shown in
Figure~\ref{fig:veldist}, with a Gaussian centred on 2903 \kms, with a
width of 190 \kms\ overlaid. The distribution of the galaxies appears
normal, with a slight skewness to higher velocities. The velocity
distribution of \HI\ detected galaxies appears the same as for all
galaxies in the region.

\begin{table} 
\centering
\caption{Characteristics of the NGC 3783 group}
\label{tab:higroup} 
\begin{tabular}{lr}
\hline
Distance [Mpc]               & 36$^{\dagger}$ \\
Luminosity Weighted Centre [$\alpha,\delta$(J2000)] & 11:37:12,-37:30:57.6 \\
Total number of galaxies      & 47\\
Number of \HI\ detections & 12\\
Mean velocity, $v$ [km s$^{-1}$]     & 2903$\pm26$  \\
Velocity dispersion, $\sigma_v$ [km s$^{-1}$]  & 190$\pm24$ \\
log $L_X$ [ergs s$^{-1}$] & 40.76$\pm0.11$$^{\dagger}$\\
%Virial Mass   [\Msun\ ]         & $6.4 \times 10^{13}$ \\
Total \HI\ mass [\Msun] & 3.7$\times 10^{10}$ \\
\hline
\end{tabular}
\flushleft{$^{\dagger}$from Osmond \& Ponman (2004)}
\end{table}

\begin{table*}
\centering
\caption{New group members discovered in the NGC 3783 Group. The
positions and velocities were derived from ATCA follow-up data, apart
from GEMS\_N3783\_10, where these values were derived from the Parkes
data. The columns are as follows: (1) GEMS source name, (2) Previous
optical ID, (3) R.A., Dec (J2000), (4) Systemic Velocity, (5) \HI\
mass derived from the Parkes data, (6) \HI\ mass derived from ATCA
data.}
\begin{tabular}{lccccccc}
\hline

GEMS Name & Optical ID &$\alpha,\delta$(J2000) & Velocity & \HI\ Mass (PKS) & \HI\ Mass (ATCA) \\
          &             &  [$^{\rm h\,m\,s}$], [\degr\,\arcmin\,\arcsec]  &  \kms    & 10$^8$\Msun     & 10$^8$\Msun     \\
(1)       &  (2)        & (3)      & (4)      & (5)             &(6)              \\
\hline
GEMS\_N3783\_2 & $\cdots$ &11:31:27,$-$36:18:37&  2731$\pm4$ & 3.8$\pm1.3$  &1.9$\pm0.3$ \\  
GEMS\_N3783\_8 & $\cdots$ & 11:38:02,$-$37:57:59 & 2983$\pm8$ & 21$\pm2.1$  & 7.0$\pm$2.0\\
GEMS\_N3783\_10& ESO 319 -G 020 & 11:26:06,$-$37:51:26 & 2810$\pm5$&7.2$\pm1.5$&$\cdots$\\
ATCA\_1134-37& $\cdots$ &11:34:02,$-$37:14:15 & 3141$\pm5$ & $\cdots$ & 2.4$\pm0.6$ &   &\\

\hline
\end{tabular}
\label{tab:newgals} 
\end{table*}

\section{High resolution observations of NGC 3783, and a new dwarf galaxy}
\label{newgal}

\begin{figure*} 
\begin{tabular}{c}
\mbox{\psfig{file=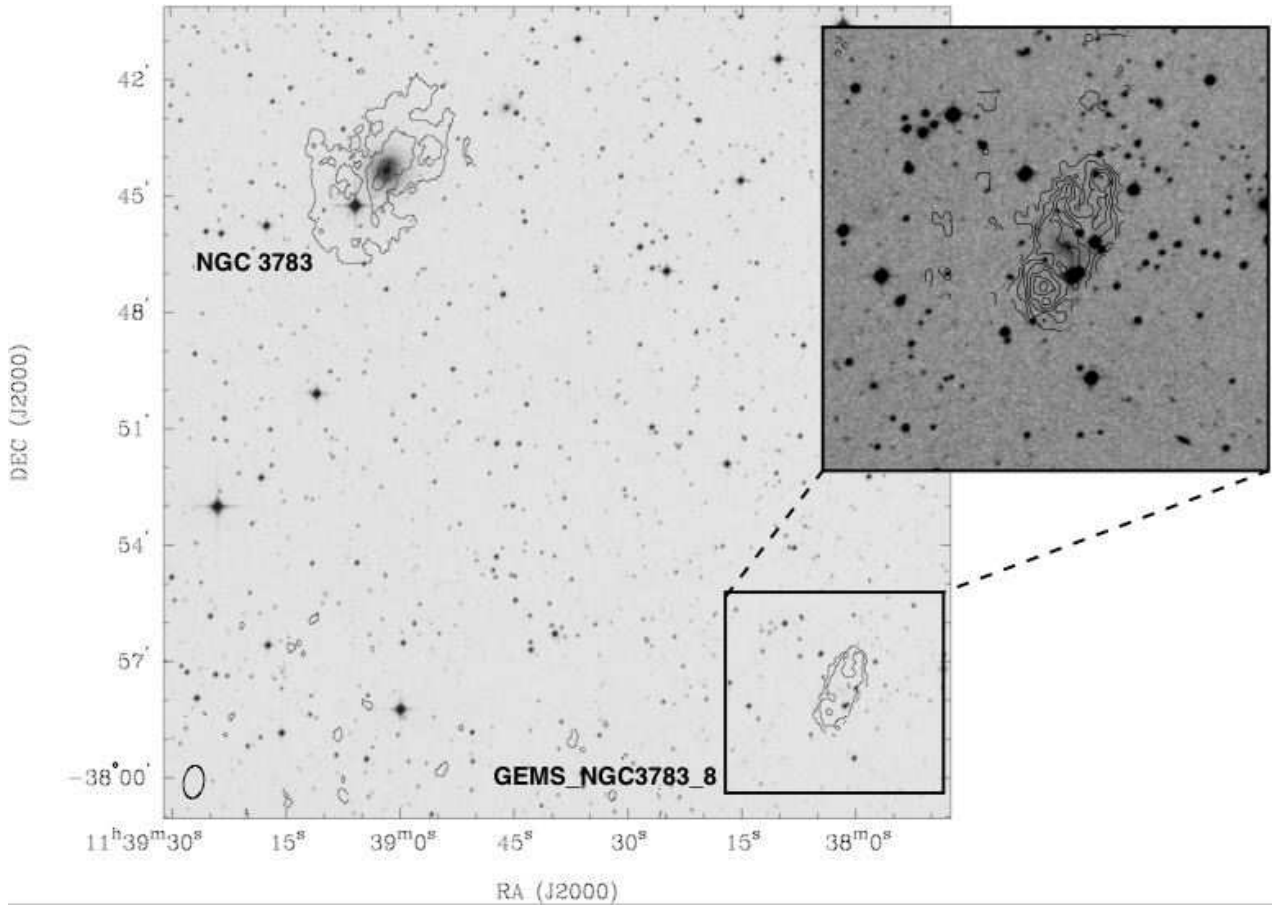,height=12cm}}
\end{tabular}
\caption{Velocity integrated \HI\ flux distribution for NGC 3783
(GEMS\_N3783\_7; top left) and GEMS\_N3783\_8 (bottom right), overlaid
on the DSS II $B$-band image. The contour levels in the main image are
0.15, 0.3, 0.6 Jy \kms. The insert is a blow-up of GEMS\_N3783\_8,
overlaid on the DSS II $B-$band image, and the contour levels are
0.15, 0.2, 0.3, 0.4, 0.5, 0.6 Jy \kms. The ATCA beam is shown in the
bottom left of the main image.}
\label{fig:newATCA}
\end{figure*}

\begin{figure*} 
\begin{tabular}{c}
\mbox{\psfig{file= 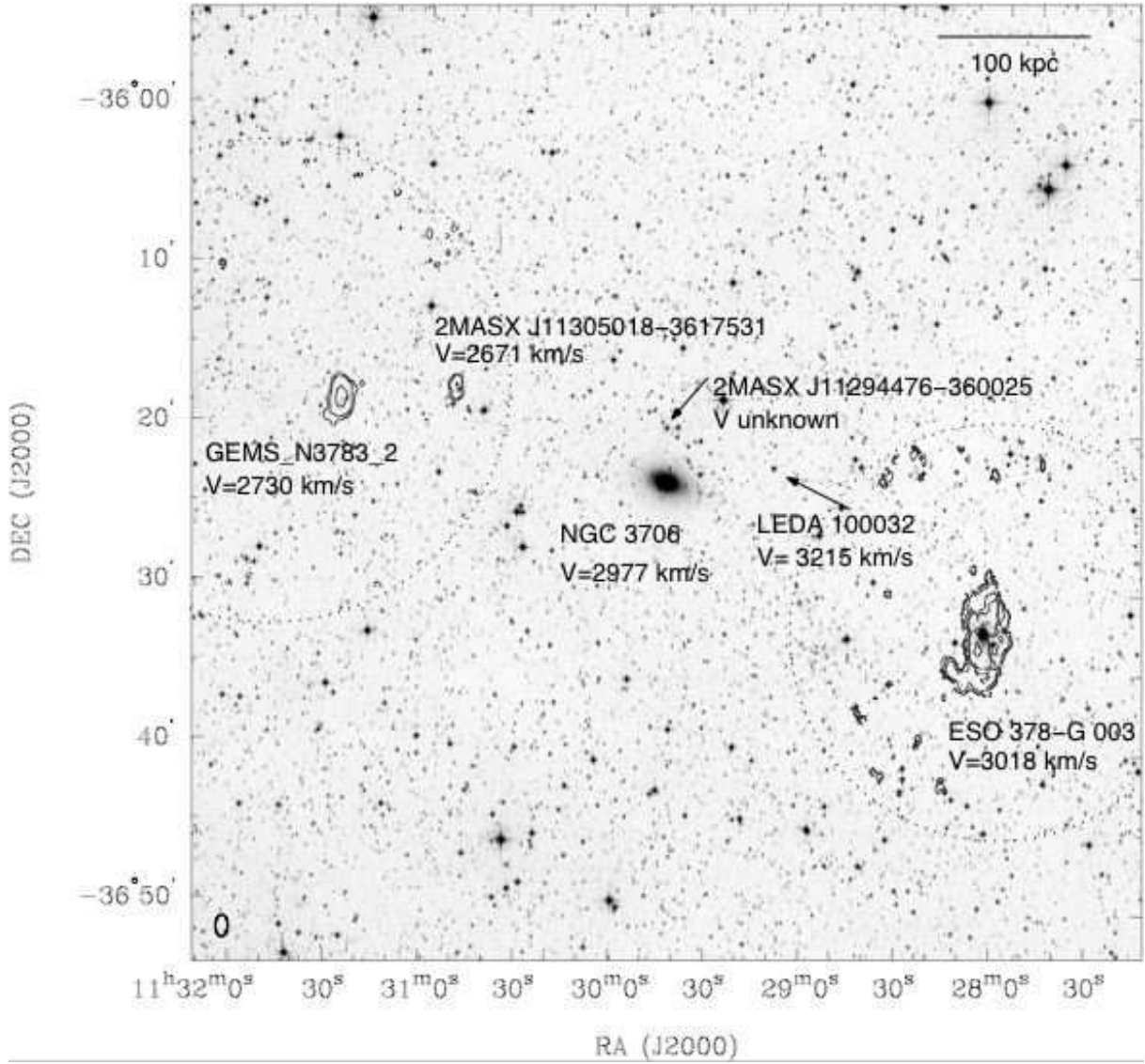,height=15cm}}
\end{tabular}
\caption{Neutral hydrogen distribution of GEMS\_N3783\_2 and ESO 378-G
003 (GEMS\_N3783\_3), overlaid on the DSS II $R$-band image. The
contours show the \HI\ emission, and the levels are 0.15, 0.2, 0.4, 0.8, 1 Jy
\kms. The ATCA beam of 88\arcsec$\times 50\arcsec$ is shown in the
bottom left corner. Two separate ATCA pointings were used, and the
noisy edge of the primary beam corrected \HI\ datacubes have been
masked out indicated by the dashed lines in the image. Along with
GEMS\_N3783\_2, we detect 2MASX J11305018-3617531 in \HI. Heliocentric
velocities are indicated.}
\label{fig:hi_cloud_region}
\end{figure*}

\begin{figure*} 
\begin{tabular}{c}
\mbox{\psfig{file=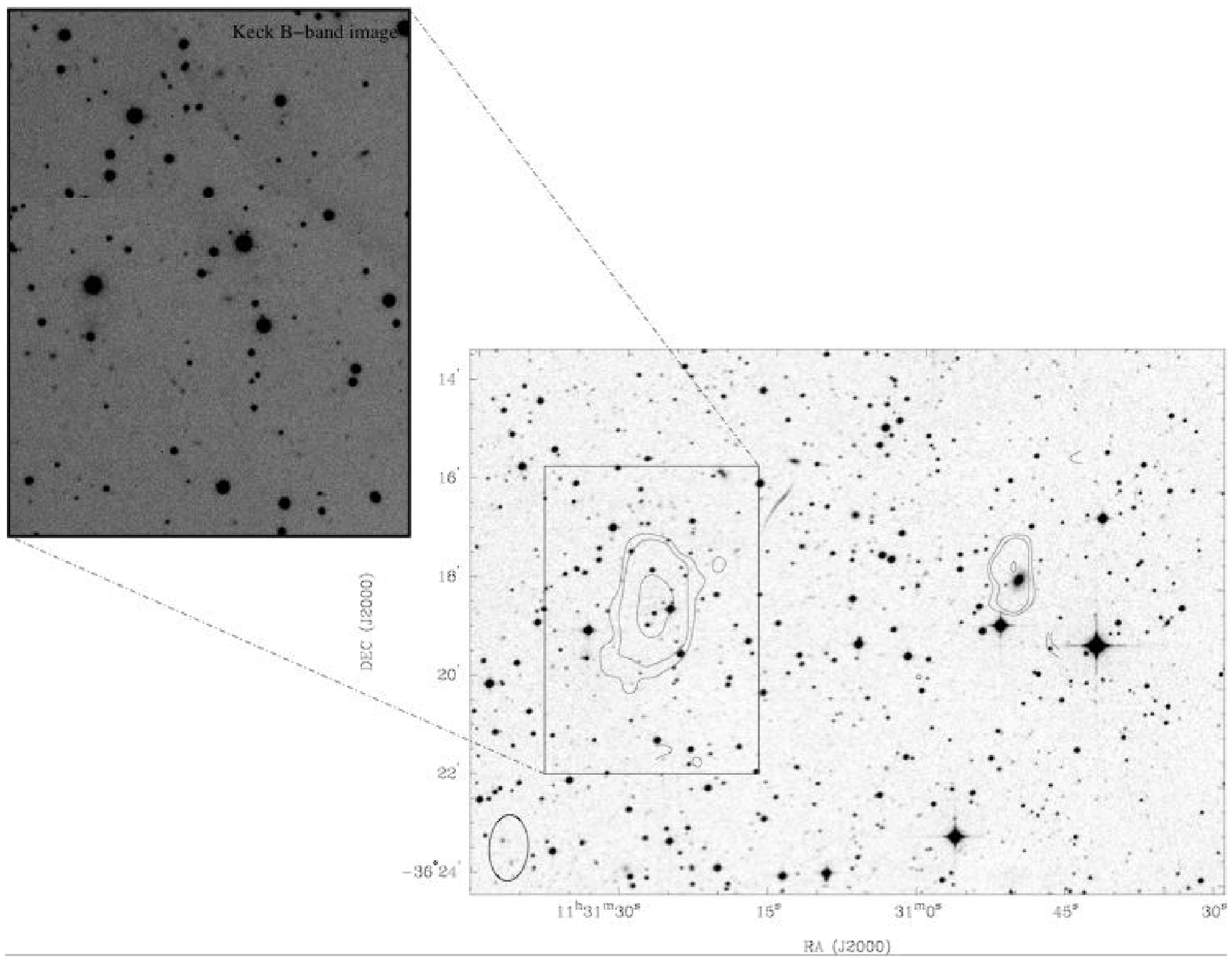,height=12cm,angle=0}}
\end{tabular}
\caption{Neutral hydrogen distribution of GEMS\_N3783\_2 (left) and
2MASX JJ11305018-3617531 (right), overlaid on the DSS II $R$-band
image. The contour levels are 0.15, 0.2, 0.3 Jy \kms. The ATCA beam of
88\arcsec$\times50\arcsec$ is shown in the bottom left corner. The
insert shows the $B$-band Keck image for the region surrounding GEMS\_N3783\_2. }
\label{fig:hi_cloud}
\end{figure*}

We obtained ATCA archive \HI\ observations of the galaxy NGC 3783 (see
Table~\ref{tab:atca} for details). Along with the detection of NGC
3783, we detected GEMS\_N3783\_8, at a distance of $\sim$15\arcmin\
($\sim$ 160 kpc) from NGC 3783, and at a velocity of 2983 \kms. The
position of the \HI\ emission from the ATCA data is
$\alpha,\delta$(J2000) = 11:38:01.8,--37:57:59, which is directly
centred on a previously uncatalogued dwarf galaxy. An ATCA \HI\ map
showing NGC~3783 and the new dwarf galaxy is shown in
Figure~\ref{fig:newATCA}. Other faint galaxies are visible in the
optical image of the observed field, but not detected in
\HI. GEMS\_N3783\_8 lies near the edge of the ATCA primary beam. For
clarity, the noisy edge of the data was masked out in this image. The
\HI\ mass for the new dwarf galaxy measured in the ATCA data is
$7\pm2\times 10^8$\Msun, compared to an \HI\ mass of 2.1$\pm0.25\times
10^9$\Msun\ from our Parkes data, thus it appears that this dwarf
galaxy has an extended \HI\ component that was resolved out in the
ATCA observations. It should be noted that as GEMS\_N3783\_8 lies near
the edge of the primary beam, the ATCA flux may be underestimated. The
ATCA position of GEMS\_N3783\_8 is indicated in
Figure~\ref{fig:hi_mom0}, and the parameters for GEMS\_N3783\_8 are
given in Table~\ref{tab:newgals}. GEMS\_N3783\_8 lies between NGC 3783
and ESO 320-G 013 both spatially (190kpc and 120 kpc projected
separation respectively), and in velocity (166 and 35 \kms\
respectively).

\section{Intra-group \HI\ Gas} 
\label{cloud}

\begin{figure} 
\begin{tabular}{c}
\mbox{\psfig{file=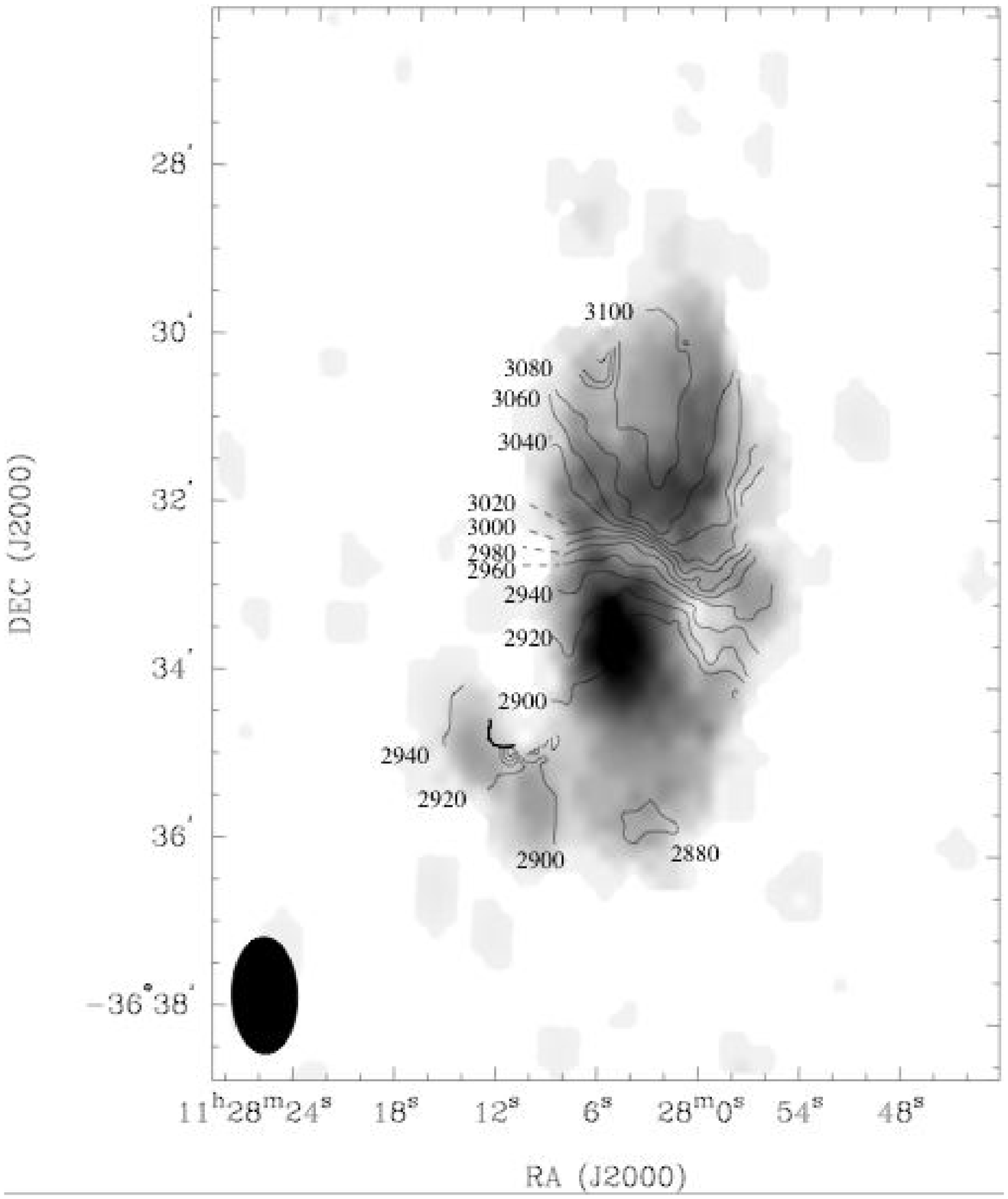,height=8cm,angle=0}}
\end{tabular}
\caption{Neutral hydrogen distribution of the spiral galaxy ESO 378-G
003 (greyscale), with velocity contours overlaid. ESO 378-G 003 shows
signs of tidal interaction in the SE part of the galaxy, with an irregular velocity field, and an extension in the \HI\ emission. }
\label{fig:n3783_3}
\end{figure}

There is one other \HI\ detection that does not correspond with any
previous NED or 6dFGS DR2 optically catalogued galaxy,
GEMS\_N3783\_2. However, unlike GEMS\_N3783\_8, inspection of optical
images obtained from the Second Generation Digital Sky Survey (DSS
II), at the position of the \HI\ detection showed no obvious
corresponding optical emission.

We obtained higher resolution follow-up observations from the ATCA for
GEMS\_N3783\_2, described in Table~\ref{tab:atca}. The ATCA velocity
integrated \HI\ flux density map overlaid on the corresponding DSS II
$R$-band image is shown in Figure~\ref{fig:hi_cloud_region} and a
close-up of the region is shown in Figure~\ref{fig:hi_cloud}. The peak
column density of GEMS\_N3783\_2 is 7.5$\times$ 10$^{19}$
cm$^{-2}$. GEMS\_N3783\_2 has a total \HI\ mass of \MHI\ =
3.8$\pm1.3$$\times 10^8$ \Msun, determined from the Parkes
observations. The ATCA \HI\ map is unresolved with the ATCA beam size
of 88\arcsec$\times 50\arcsec$. The \HI\ parameters for GEMS\_N3783\_2
are given in Table~\ref{tab:cloud}. The \HI\ mass detected by the ATCA
is 1.9$\pm0.3\times 10^8$ \Msun, slightly lower than the Parkes
observations, indicating we are missing some extended \HI\
emission. We made optical observations in the region of this object
with the Keck telescope on 2005 Feb. 9 (see
Figure~\ref{fig:hi_cloud}). Images in $R$ and $B$-bands do not show
the presence of any low surface brightness galaxy down to a limiting
surface brightness of $B \sim$ 22 mag arcsec$^{-2}$.

There are several small, faint optical sources within the \HI\
emission region for GEMS\_N3783\_2. We obtained a short service time
observation with the 2dF multi-fibre spectrograph on 2005 May 4 to
obtain redshifts for nearby bright, resolved optical sources. The 2dF
spectrograph covers a 2 degree field which could encompass the region
about GEMS\_N3783\_2 in the one pointing. Most of the sources we
observed did not contain emission or absorption lines for us to
determine a redshift from. No sources were found at the group
velocity.

\begin{figure*} 
\begin{tabular}{c}
  \mbox{\psfig{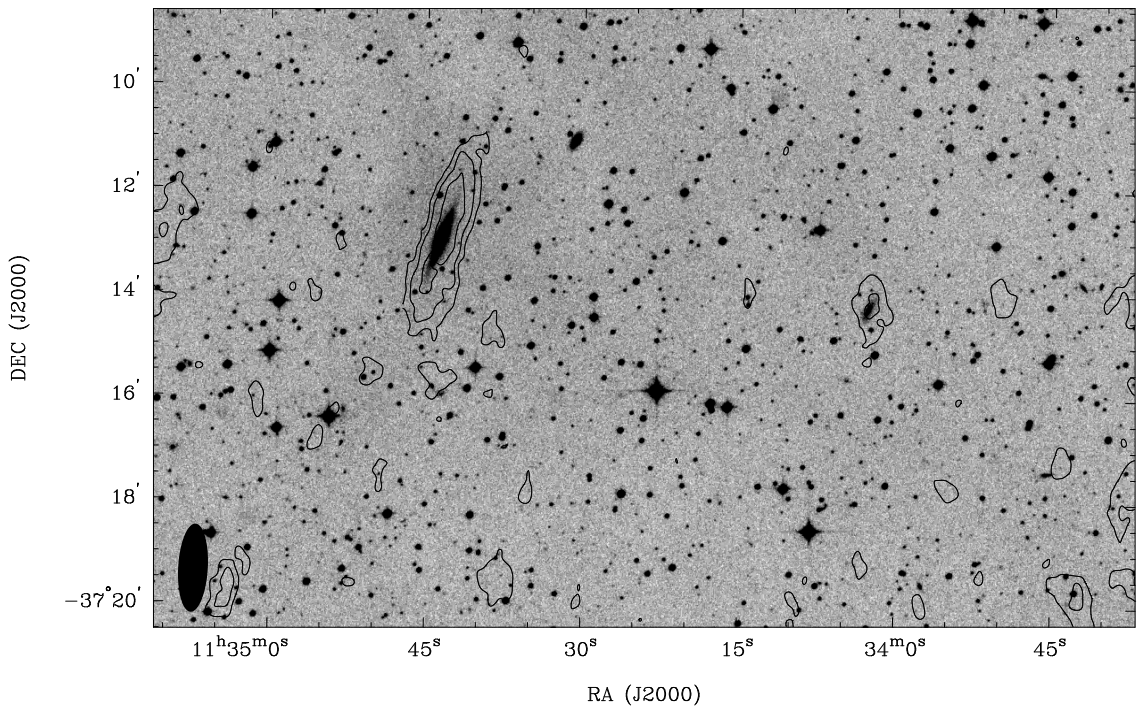}}
\end{tabular}
\caption{Neutral hydrogen distribution of ESO 378-G 011 (left), and
the small, previously uncatalogued dwarf galaxy ATCA\_1134-37 (right),
overlaid on the DSS II $R$-band image. The contour levels are 0.5, 1,
2, 3 Jy \kms.}
\label{fig:n3783_1}
\end{figure*}

What is the origin of GEMS\_N3783\_2? We have little information as
yet on the structure or velocity field of this \HI\ region, making it
hard to determine the origin by studying it. Examining
Figure~\ref{fig:hi_cloud_region}, The nearest bright galaxy to
GEM\_N3783\_2 is the early-type galaxy NGC 3706. There are several
nearby dwarf galaxies, and the nearest gas-rich spiral galaxies are $>
500$ kpc away.  We have obtained high resolution ATCA \HI\
observations of the nearest spiral galaxies. The \HI\ distribution
of ESO 378-G 003 is shown in Figure~\ref{fig:hi_cloud_region} and
Figure~\ref{fig:n3783_3}, and the \HI\ distribution of ESO 378-G 011
is shown in Figure~\ref{fig:n3783_1}.

ESO 378-G 003 lies at a projected distance of $\sim 235$ kpc from NGC
3706, and $\sim 450$ kpc from GEMS\_N3783\_2. ESO 378-G 003 shows an
extended \HI\ distribution, which is distorted on the SE side of the
galaxy. The velocity contours shown in Figure~\ref{fig:n3783_3} show
typical spiral rotation in the undisturbed side of the galaxy, while
the extended part shows irregular rotation, indicating a tidal
interaction. 

Figure~\ref{fig:n3783_1} shows the \HI\ distribution for ESO 378-G 011
overlaid on the DSS II $R$-band image. In our ATCA \HI\ datacube, we
also detect a previously uncatalogued dwarf galaxy, ATCA\_1134-37, at
the position $\alpha,\delta$(J2000) = 11:34:03,$-$37:14:22).  ESO 378-G
011 lies at a projected distance of $\sim$ 570 kpc from
GEMS\_N3783\_2, and the \HI\ distribution for the galaxy appears
regular.

%If GEMS\_N3783\_2 was formed through a tidal interaction of a spiral
%galaxy in the past, then the nearest spiral galaxies (ESO 378-G 011
%and ESO 378-G 003) may contain less \HI\ than expected for their
%optical morphology and size (e.g. Kilborn et al. 2005; Haynes \&
%Giovanelli 1984). However, the expected \HI\ for a spiral galaxy of
%the diameters of ESO 378-G 011 and ESO 378-G 003 (following Kilborn et
%al. 2005) are both $\sim 3\times10^9$ \Msun, which is lower than the
%actual detected amount of \HI\ for these galaxies; $1.3 \times
%10^{10}$ \Msun\ and $5.2\times10^9$ \Msun, respectively. These spiral
%galaxies look undisturbed in the optical.

\begin{table*} 
\centering
\caption{Neutral hydrogen parameters of GEMS\_N3783\_2.}
\label{tab:cloud} 
\begin{tabular}{lcc}
\hline
& Parkes & ATCA\\
\hline
Central Position [$\alpha,\delta$ (J2000)] & 11:31:32, -36:18:53  &11:31:27, -36:18:37 \\
\HI\ Mass [$10^8$ \Msun] & 3.8$\pm1.3$  &1.9$\pm0.3$ \\
Velocity [km s$^{-1}$] & 2730 & 2731\\
$\Delta v_{20}$ [km s$^{-1}$] & 116 & 50\\
$\Delta v_{50}$ [km s$^{-1}$] & 106 & 40 \\
\hline
\end{tabular}
\flushleft
\end{table*}

\section{Discussion}

The NGC 3783 group is a loose group of galaxies, with diffuse X-ray
emission centred on the galaxy NGC 3783 itself. What is the
evolutionary state of this group -- for example, is it virialised? The
extent of the X-ray emission is 69 kpc, which is low for intra-group
X-ray emission, and only a single component fit to the surface
brightness profile of the X-ray emission was possible. For comparison,
Mulchaey \& Zabludoff (1998) find that for nine X-ray detected groups
it was possible to make a two component fit to the surface brightness
of the X-ray emission, with the first component extending 20-40 kpc,
and corresponding to emission from the galaxy, and the second
component extending 100-300 kpc, and corresponding to diffuse
intra-group X-ray emission. The extent of the X-ray emission in the
NGC~3783 group could be interpreted in two ways: The X-ray emission
could emanate solely from the Seyfert galaxy NGC~3783 itself, or the
X-ray emission could be from the combination of emission from
NGC~3783, and the group potential of a newly forming galaxy group,
perhaps one that is not yet virialised. Looking at the velocity
distribution of the group, it is nearly normally distributed, usually
a sign of a relaxed system of galaxies. However, the velocity-distance
plot shown in Figure~\ref{fig:veldist} does not look like a typical
virialised group. The galaxy NGC 3783 itself is the closest galaxy to
the luminosity weighted centre of the group, although it is 267 kpc
from this centre, which is unusual for a virialised group (Brough et
al. 2006). NGC 3783 lies at the mean velocity of the group. This group
seems to display some characteristics of a virialised group, however
the fact that the X-ray emission is centred on a late-type galaxy, and
is offset from the centre of the group indicates that this may be an
example of a group in the early stages of evolution.

In our neutral hydrogen survey of the NGC 3783 group, we found several
new group members, and one region of \HI\ emission that appears to
have no stars associated with it. While there are many examples of
isolated neutral hydrogen that has been removed from galaxies (e.g. HI
rogues gallery (Hibbard et al. 2001a):
http://www.nrao.edu/astrores/HIrogues/), this is one of the most
extreme examples of isolated \HI\ ever found. To date, all neutral
hydrogen that has been found in emission can be associated with either
galaxies, regions of stars or star formation.

There are three possible explanations for the existence of
GEMS\_N3783\_2 in the NGC 3783 galaxy group. Either this is an
extremely low surface brightness galaxy that we have been unable to
detect optically yet,  GEMS\_N3783\_2 is comprised of gas that has
been removed from a galaxy in the past, or GEMS\_N3783\_2 is a
primordial cloud of neutral hydrogen which has not formed stars.

If GEMS\_N3783\_2 is a low surface brightness galaxy, then the limits
placed by our Keck observations are such that the central surface
brightness must be less then 22 mag arcsec$^{-2}$ in the
$B$-band. Assuming a source size of about 10\arcsec\ (e.g. the size of
the nearest dwarf galaxy to GEMS\_N3783\_2, 2MASX J11305018-3617531),
then we would have detected a typical dwarf galaxy in our optical
imaging. However, we are not sensitive to an extended low surface
brightness galaxy, and deeper optical imaging is needed to confirm
there is no stellar component to GEMS\_N3783\_2.

GEMS\_N3783\_2 may be a remnant of stripping of a gas-rich
galaxy. It is unlikely that GEMS\_N3783\_2 was formed through
ram pressure stripping, as the hot gas in the NGC 3783 group is
confined to the region around NGC 3783 itself, and that the projected
distance of GEMS\_N3783\_2 from the X-ray emission is $>500$
kpc. GEMS\_N3783\_2 may have formed through a tidal interaction
between a gas-rich galaxy, and one or more other galaxies in the
group.

The nearest gas-rich galaxy to GEMS\_N3783\_2 is 2MASX
J11305018-3617531, for which we detected an \HI\ mass of $\sim 10^7$
\Msun\ in our ATCA observations. Given the \HI\ mass of GEMS\_N3783\_2
at $\sim4\times10^8$ \Msun\ is typical of a dwarf irregular galaxy
(eg: Salzer et al. 2002; Stil \& Israel 2002; Hoffman et al. 1996),
perhaps 2MASX J11305018-3617531 has been tidally influenced by NGC
3706, removing the majority of the \HI\ from the galaxy. This seems
unlikely as we do not detect a bridge of \HI\ joining GEMS\_N3783\_2 and
2MASX J11305018-3617531, and to remove the majority of \HI\ from a
dwarf galaxy while leaving the stars intact would require an
extremely extended \HI\ distribution of the dwarf galaxy.

The nearest gas-rich spirals to GEMS\_N3783\_2 are ESO 378-G 003 and
ESO 378-G 011, at a projected separation of $\sim 450$ and $\sim 570$
kpc respectively. Given the large projected separation, and apparently
undisturbed \HI\ distribution of ESO 378-G 011, we do not consider
that it is the origin of GEMS\_N3783\_2. On the other hand, ESO 378-G
003 is closer to GEMS\_N3783\_2, and importantly, is also very close
to the bright early-type galaxy NGC 3706. ESO 378-G 003 displays
evidence of tidal interaction in its irregular \HI\ and velocity
distribution on one side of the galaxy.  If GEMS\_N3783\_2 was formed
from the interaction of NGC 3706 and ESO 378-G 003, then using the
difference in velocity of GEMS\_N3783\_2 and ESO 378-G 003, and the
projected separation of 450 kpc, the time-scale for the interaction is
$\sim 1.5$ Gyr. Of the many possibilities, this seems the most likely,
but it requires deeper observations to confirm. Deeper observations
might uncover further \HI\ in the system, as in the case of the Leo
cloud (Schneider 1985), and the NGC 1490 system (Oosterloo et al. 2004).

Finally, could GEMS\_N3783\_2 be an isolated primordial \HI\ cloud,
that has lived quiescently forming no stars? Cosmological simulations
of groups have predicted many more dark matter halos than optical
galaxies are seen observationally (eg: D'Onghia \& Lake 2004; Moore et
al. 1999; Klypin et al. 1999). Perhaps GEMS\_N3783\_2 is an example of
such a dark halo, containing primordial \HI\ gas? While not
impossible, this option does seem unlikely.  Of our \HI\ survey of 16
galaxy groups, GEMS\_N3783\_2 is the only detection that we have found
which does not have an obvious galaxy associated with it. If
GEMS\_N3783\_2 is primordial material, then it is the only example we
have found of such an object in our \HI\ survey, and thus the number
of these objects appears to be very low. We note, that our survey of
16 groups has a mass limit of $\gtrsim 5 \times 10^8$ \Msun, so we are
unable to say anything about the population of lower mass \HI\ clouds,
however other authors have not found any low mass \HI\ clouds in deep
\HI\ studies of loose groups (Pisano 2004; Zwaan 2001).

\section{Conclusions}

We have made a wide-field \HI\ survey of the NGC~3783 galaxy group,
using the Parkes radiotelescope. We found 12 \HI\ detections in the
region, of which one is a region of extended \HI\ emission that we
cannot match with corresponding optical emission. We believe the
origin of this \HI\ region is likely to be tidal debris rather than
ram-pressure stripped, or primordial \HI. We found two previously
uncatalogued dwarf galaxies -- one from our Parkes observations
(GEMS\_N3783\_8), and one from high resolution ATCA observations
(ATCA\_1134-37).  We used the 6dFGS DR2 and NED databases to find
previously catalogued galaxies in the region, and determine parameters
for the NGC~3783 group. We calculate a mean velocity for the group of
2903$\pm$26 \kms, and velocity dispersion of 190$\pm$24 \kms. The
NGC~3783 group is a rare example of a group at the early stage of its
evolution. Comparison with other GEMS groups will help in determining
the evolutionary path of galaxies within groups in general.

\section*{Acknowledgments}

We thank David Barnes for invaluable help in reducing the Parkes
narrow-band data, and for making modifications to the reduction
software.  We thank Rob Sharp for taking the 2dF observations of
GEMS\_N3783\_2 optical candidates. We acknowledge Heath Jones and
Matthew Colless for providing helpful information on the 6dFGS
data. We thank Jean Brodie for the Keck observations of
GEMS\_N3783\_2. John Osmond and Trevor Ponman are acknowledged for
helping us with the X-ray data. We thank John Reynolds for support
with the Parkes observations. We thank the anonymous referee for useful
comments.

This research has made use of the NASA/IPAC Extragalactic Database
(NED) which is operated by the Jet Propulsion Laboratory, California
Institute of Technology, under contract with the National Aeronautics
and Space Administration. This publication makes use of the Two Micron
All Sky Survey (2MASS) which is a joint project of the University of
Massachusetts and the Infrared Processing and Analysis
Centre/California Institute of Technology, funded by the National
Aeronautics and Space Administration and the National Science
Foundation. The Digitized Sky Survey was produced at the Space
Telescope Science Institute under U.S. Government grant NAG W-2166.

\label{lastpage}

\end{document}